# An RBF-based method for computational electromagnetics with reduced numerical dispersion

Andrej Kolar-Požun [a,b,*], Gregor Kosec [a]

[a] Parallel and Distributed Systems Laboratory, Jožef Stefan Institute, Jamova 39, 1000, Ljubljana, Slovenia
[b] Faculty of Mathematics and Physics, University of Ljubljana, Jadranska 19, 1000, Ljubljana, Slovenia

ARTICLE INFO

ABSTRACT

The finite difference time domain method is one of the simplest and most popular methods in computational electromagnetics. This work considers two possible ways of generalising it to a meshless setting by employing local radial basis function interpolation. The resulting methods remain fully explicit and are convergent if properly chosen hyperviscosity terms are added to the update equations. We demonstrate that increasing the stencil size of the approximation has a desirable effect on numerical dispersion. Furthermore, our proposed methods can exhibit a decreased dispersion anisotropy compared to the finite difference time domain method.

## 1. Introduction

Accurate simulations of electromagnetic waves have been of great interest lately with one of the key drivers being the field of telecommunications, as the carrier frequencies for future communication protocols (e.g. 6G) are expected to be higher, demanding physically accurate simulations that are able to describe increasingly finer geometrical objects, such as antennas. In order to satisfy that requirement, we must solve the equations that describe the physics of electromagnetism – Maxwell's equations. Maxwell's equations form a coupled system of first order Partial Differential Equations (PDEs) that are too complex to be solved analytically in realistic setups. In order to obtain the solution, a numerical procedure must be invoked.

Most of the commonly used methods in computational electromagnetics have been around for decades with several landmark papers appearing in the 1960s. In 1966, for instance, the well known Finite Difference Time Domain (FDTD) method was proposed by Yee [1]. In the same year, the Method of Moments (MoM) was popularised by Richmond [2]. The Finite Element Method (FEM), which can be traced all the way to the work of Courant in 1943 [3], has also been applied to electromagnetics by Silvester in 1969 [4]. A comprehensive review and comparison of these can be found in any of the major textbooks on computational electromagnetics, such as [5]. Among the listed methods, the FDTD is arguably the simplest to implement and therefore a common starting point when dealing with a given problem. Its simplicity lies in the fact that the FDTD is, in essence, just the application of the usual Finite Difference Method (FDM) on a staggered grid. This approach, however, has its limitations – it is limited to regular, grid-like geometries, making it difficult to efficiently satisfy our requirement of describing fine geometrical objects. Additionally, the underlying grid makes the numerical dispersion relation anisotropic, even if the propagation medium is isotropic.

* Corresponding author.
*E-mail addresses:* andrej.pozun@ijs.si (A. Kolar-Požun), gregor.kosec@ijs.si (G. Kosec).

Our goal in this paper is to remedy both of these shortcomings of the FDTD by introducing a method that in a certain sense generalises it to irregular node layouts, providing a greatly increased geometrical flexibility.

For that purpose, we will adopt meshless methods, which provide the means of function approximation on scattered nodes without any additional structure. The terms “meshless” or “meshfree” encompass a variety of different methods, some of which are described in, for example [6]. For our work we opt for methods based on Radial Basis Functions (RBFs) [7]. RBFs first appeared in the context of scattered data interpolation in the 1970s [8] and were applied to differential operator approximation in the 1990s [9]. In the 2000s a local RBF-based operator approximation method was introduced [10], which directly generalises the Finite Difference Method (FDM), sparking interest from the wider scientific community and resulting in a culmination of research in the field, both theoretical and experimental [11,12].

With the machinery of meshless methods on hand, a generalisation of the FDTD is then straightforward – we can simply replace the Finite Differences (FD) appearing in the FDTD by their appropriate meshless variants. We will see that there are two natural ways of doing so, resulting in two different generalisations to consider.

At this point a natural question may arise - why should one bother with a meshless variant of the FDTD when FEM already solves Maxwell’s equations (albeit in weak form) and is known to be geometrically flexible? Our motivation for considering meshless methods over FEM is twofold. Firstly, unlike meshless methods, FEM requires an underlying mesh, a generation of which can be tedious and cannot be fully automated, especially in higher dimensions [13]. Secondly, FEM is a fairly complex method – the attractive point of the FDTD is its simplicity and our aim is to preserve this property in the geometrically flexible version. Simplicity of the method can be of great importance when dealing with problems of increasing complexity – not only with regards to the problem geometry but also, for example, inclusion of non-homogenous or even non-linear materials. Another aspect of great practical value is parallelisation, which is easy to implement in an explicit method, such as the FDTD, and generally difficult in FEM [5].

It should be mentioned that computational electromagnetics has already been studied in meshless context. Much of existing work has been focused on weak-form meshless methods [14–16], although some research has been done also using the Method of Fundamental Solution (MFS) [17], which is closely related to the previously mentioned method of moments. However, our interests lie in strong-form methods, just like the FDTD. Existing work has been done also in this case [18–20], where different authors adopted a similar idea of replacing the spatial derivatives appearing in the FDTD by their meshless variants in a similar way as we intend to. However, our proposed methods differ from all of the listed work by two key points. Firstly, previous authors have always staggered the $\vec{E}$ and $\vec{H}$ fields on different nodes, as in the usual FDTD. This staggering can become non-trivial in a scattered setting and would require, for example, a Voronoi tessellation, which some would argue is not truly a meshless method. The second key point is perhaps more important – despite the methods being meshless in their formulations, they seemed to have only been tested on regular node layouts, where the methods can be much better behaved, particularly in terms of stability. In our approach we abandon the staggering of the $\vec{E}$ and $\vec{H}$ fields and our main goal is to demonstrate that our methods work also on scattered, irregular nodes.

The paper is structured as follows: In Section 2, the governing equations of electromagnetism are listed. In Section 3, the FDTD is briefly described, along with its previously mentioned shortcomings. In Section 4, RBF interpolation-based meshless methods are introduced. In Section 5, our proposed methods are introduced. In Section 6, we analyse the numerical dispersion of our proposed methods and compare it to the FDTD. A final comparison with FDTD is done on a simple scattering scenario in Section 7 followed by a conclusion in Section 8.

## 2. Governing equations

In electromagnetics, the relevant quantities are the electric and magnetic fields, which we denote by $\vec{E}$ and $\vec{H}$, respectively. They are defined on some region of interest $\Omega \subset \mathbb{R}^3$ and are generally time-dependent:

$$\vec{E}, \vec{H} : \Omega \times \mathbb{R}_{\geq 0} \to \mathbb{R}^3. \tag{1}$$

The two fields satisfy Maxwell’s curl equations, which in vacuum read:

$$\begin{aligned} \frac{\partial \vec{H}}{\partial t} &= -\frac{1}{\mu_0} \nabla \times \vec{E}, \\ \frac{\partial \vec{E}}{\partial t} &= \frac{1}{\epsilon_0} \nabla \times \vec{H}, \end{aligned} \tag{2}$$

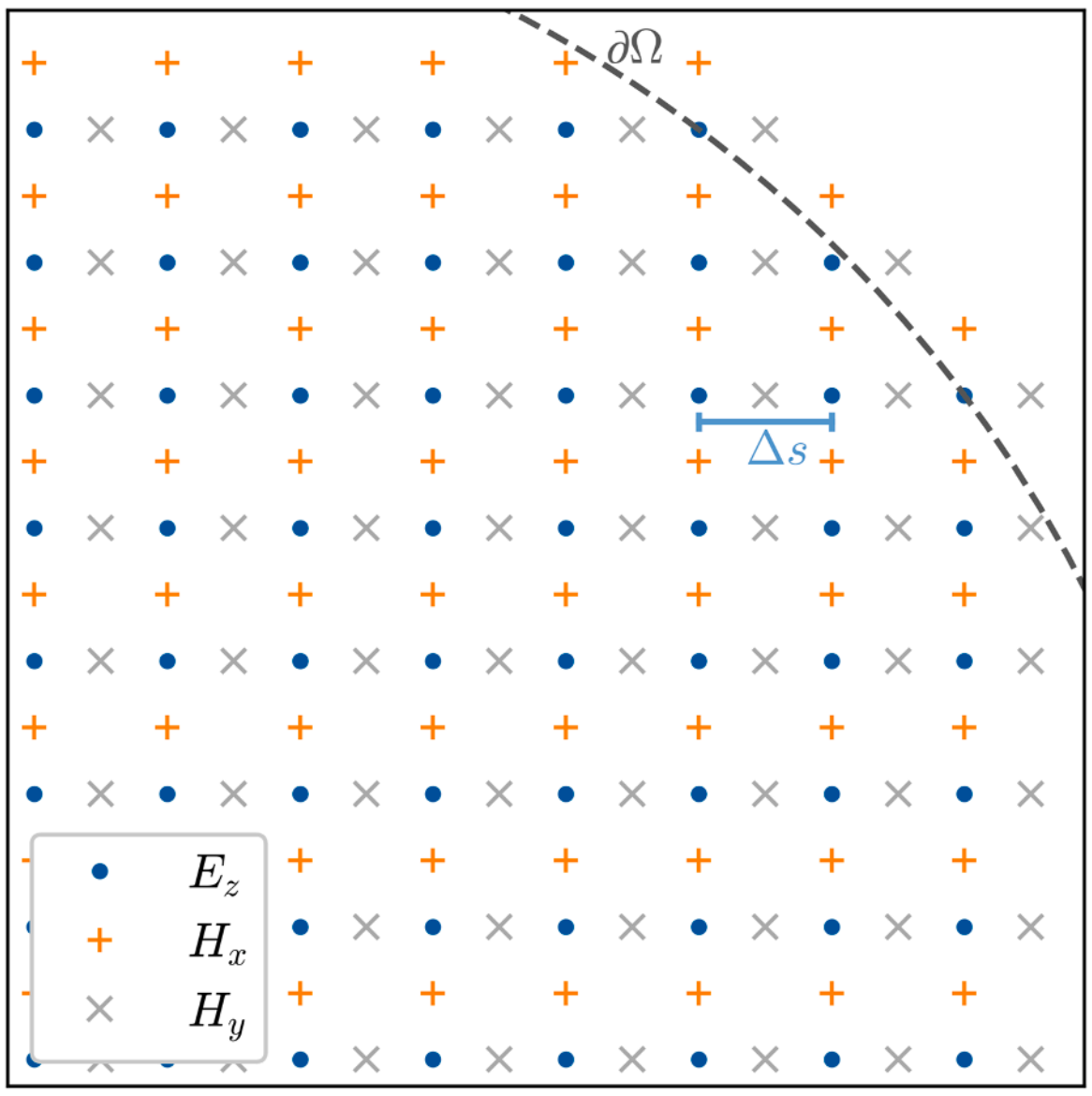

**Fig. 1.** Yee grid with a spacing of $\Delta s$. An example domain boundary $\partial\Omega$ is also displayed.

where $\epsilon_0$ and $\mu_0$ are the vacuum permittivity and permeability, respectively. Forthcoming analyses will be performed on an effectively 2-dimensional setup,[1] where the fields $\vec{E}$ and $\vec{H}$ only depend on $(x, y)$ coordinates. For this case, it turns out [21] that the above system decouples in two systems, corresponding to the so-called TMz and TEz modes – transverse magnetic and electric, respectively, in the $z$ direction.

TMz mode considers $\vec{E}(x, y) = (0, 0, E_z(x, y))$ and $\vec{H} = (H_x(x, y), H_y(x, y), 0)$, for which the equations simplify to:

$$\begin{aligned} \frac{\partial H_x}{\partial t} &= -\frac{1}{\mu_0}\frac{\partial E_z}{\partial y}, \\ \frac{\partial H_y}{\partial t} &= \frac{1}{\mu_0}\frac{\partial E_z}{\partial x}, \\ \frac{\partial E_z}{\partial t} &= \frac{1}{\epsilon_0}\left(\frac{\partial H_y}{\partial x} - \frac{\partial H_x}{\partial y}\right), \end{aligned} \tag{3}$$

and analogously for the TEz mode with the roles of $\vec{E}$ and $\vec{H}$ interchanged. For that reason we will focus only on the TMz mode during our studies.

## 3. The finite difference time domain

One of the simplest and most widespread algorithms for solving the aforementioned equations is the Finite Difference Time Domain (FDTD) method, also known as the Yee algorithm. As the name implies, the algorithm discretises the curl equations using the usual Finite Difference (FD) schemes. What sets FDTD algorithm apart from the Finite Difference Method (FDM) is the use of a staggered grid (in this context known as a Yee grid) – different components of $\vec{E}$ and $\vec{H}$ fields are discretised at different points in time and space. In the case of a TMz mode the field values are discretised as follows:

$$\begin{aligned} E_z^{n,i,j} &:= E_z(x = i\Delta s,\ y = j\Delta s,\ t = n\Delta t), \\ H_x^{n,i,j} &:= H_x(x = i\Delta s,\ y = (j+0.5)\Delta s,\ t = (n-0.5)\Delta t), \\ H_y^{n,i,j} &:= H_y(x = (i+0.5)\Delta s,\ y = j\Delta s,\ t = (n-0.5)\Delta t, \end{aligned} \tag{4}$$

which is displayed on Fig. 1.

[1] The proposed methods admit a straightforward generalisation to 3D.

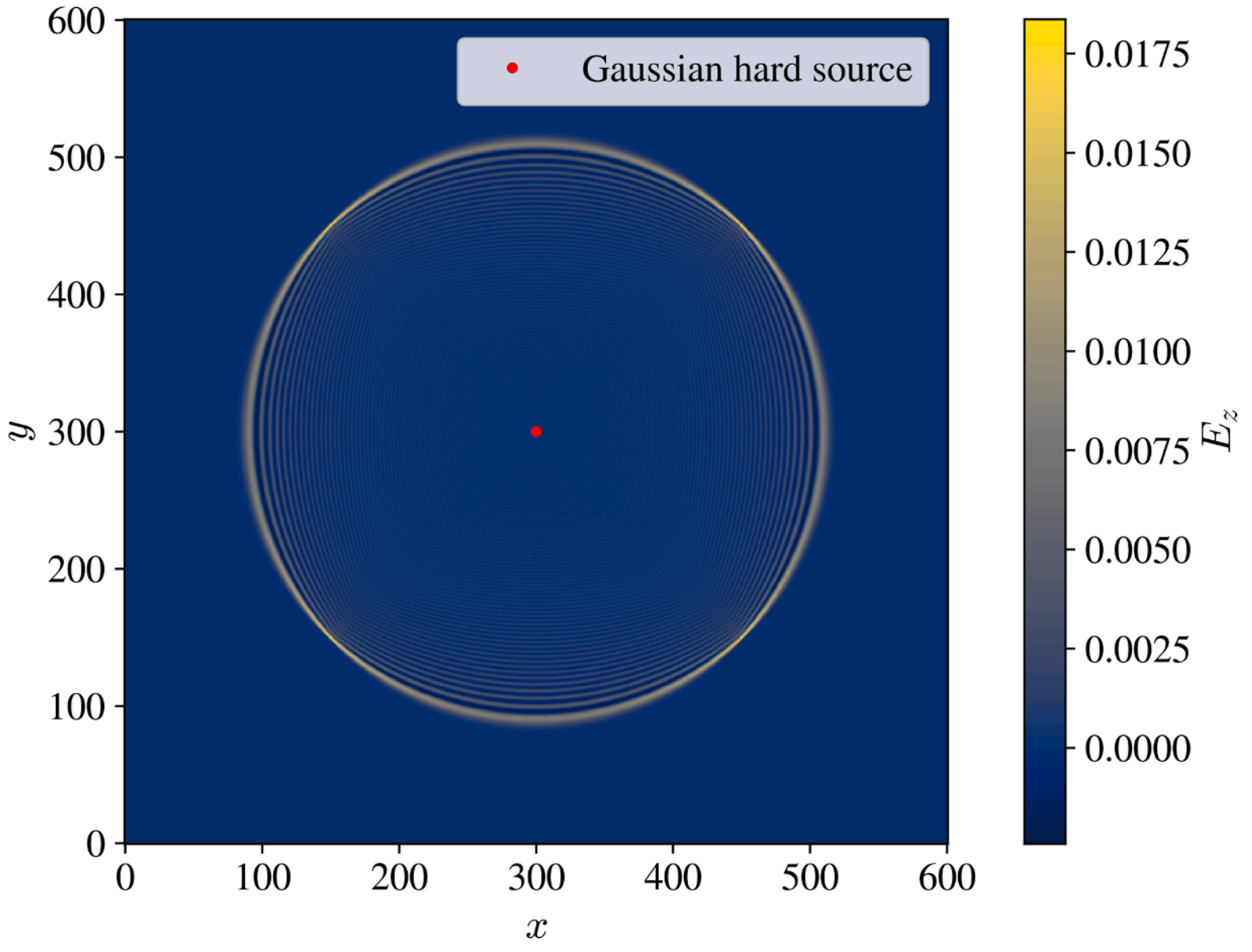

**Fig. 2.** Snapshot of the FDTD simulation at $n = 400$, demonstrating anisotropic dispersion relation.

Due to the layout of the staggered grid, all of the derivatives arising in a system of Eq. (3) can then be discretised using central differences, resulting in an explicit scheme:[2]

$$\begin{aligned} H_x^{n+1,i,j} &= H_x^{n,i,j} - S_c/\eta_0\left(E_z^{n,i,j+1} - E_z^{n,i,j}\right) \\ H_y^{n+1,i,j} &= H_y^{n,i,j} + S_c/\eta_0\left(E_z^{n,i+1,j} - E_z^{n,i,j}\right) \\ E_z^{n+1,i,j} &= E_z^{n,i,j} + \eta_0 S_c\left(H_y^{n+1,i,j} - H_y^{n+1,i-1,j} - H_x^{n+1,i,j} + H_x^{n+1,i,j-1}\right), \end{aligned} \tag{5}$$

where $S_c = \frac{c_0 \Delta s}{\Delta t}$ is the Courant number, $c_0^2 = (\epsilon_0\mu_0)^{-1}$ the speed of light, and $\eta_0^2 = \frac{\mu_0}{\epsilon_0}$ the vacuum impedance. It can be shown that the above scheme is stable if $S_c \leq \frac{\sqrt{2}}{2}$. In practice, simulations are performed at the upper bound of $S_c = \frac{\sqrt{2}}{2}$ – it turns out that this value of the Courant number minimises numerical dispersion. For further details on the algorithm and the theory behind it we refer the reader to any FDTD textbook, such as [21].

### 3.1. Shortcomings of FDTD

The FDTD has two main shortcomings that we aim to address with the proposed methods. The first could already be observed in Fig. 1 – a grid-based method has trouble describing irregular boundaries, and the domain has to terminate in a staircase manner. Such a poor boundary description can greatly affect accuracy in more realistic geometries and often, the grid is forced to be made increasingly finer, resulting in very long computational times. Remedies for that, such as the contour FDTD exist [5], but they require manual intervention and can become cumbersome for increasingly complex domains.

The second shortcoming can be seen on Fig. 2, where we simulate the effect of a hard[3] Gaussian point source, $E_z(n) \propto e^{-(n-100)^2/2}$, where $n$ is the timestep. We can see numerical dispersion appear in the form of additional, unphysical waves behind the main wave front. The interesting part is the fact that this dispersion is angle dependent – with more dispersion in the $x, y$ directions and less on the $y = \pm x$ diagonals.

In fact, FDTD numerical dispersion relation can be derived [21] and equals

$$\left(\frac{\Delta s}{c\Delta t}\right)^2 \sin^2\left(\frac{\omega \Delta t}{2}\right) = \sin^2\left(\frac{\Delta s k_x}{2}\right) + \sin^2\left(\frac{\Delta s k_y}{2}\right), \tag{6}$$

where the wave vector is $\vec{k} = (k_x, k_y) = k(\cos\phi, \sin\phi)$. As observed, the dispersion relation is anisotropic and depends on the angle of propagation $\phi$, even though the underlying physics is isotropic – a physically correct dispersion relation is well-known: $\omega = c\|\vec{k}\|$.

[2] This manner of time discretisation is also known as interleaved leapfrog integration.
[3] Hard source refers to setting the field values to a desired source function at each timestep.

The cause of this unphysical anisotropy is the presence of a grid, which breaks the symmetry in a sense that the $x$ and $y$ directions are special, despite the fact that the curl operator appearing in Maxwell's equations is rotationally symmetric.

It should be mentioned that an adequately dense grid can greatly reduce the effects of numerical dispersion. However, increasing the grid density also increases the computational time of the simulation and may not be viable. In that case, existing methods of reducing numerical dispersion are available, such as by introducing artificial anisotropy [22], working with nonorthogonal grids [23] or simply by using higher order FDTD [24]. Reducing numerical dispersion by increasing the order of the method has been shown to work also in other numerical schemes, such as in weak form methods [25]. As a remark, we should mention that in our proposed methods increasing the order should also be straightforward, although we will not investigate this further. Instead, our goal is to show that our proposed meshless generalisations of the FDTD can reduce numerical dispersion (or at least the extent of its angle dependence) on their own without any major modifications.

## 4. Computation on scattered nodes

To improve the flexibility of the numerical method and ease the description of irregular boundaries, we abandon the idea of a grid and allow the nodes to be positioned irregularly, such as in Fig. 3. To obtain such a set of nodes we use a scattered node generation algorithm, described in [26], which efficiently discretises a given computational domain, producing nodes with the internodal spacing of approximately $h$ (a chosen parameter). We will denote the nodes of the resulting discretisation as $\mathbf{x_i} = (x_i, y_i)$, where $i$ ranges from 1 to $N$, the discretisation size.

### 4.1. *Interpolation on scattered nodes*

In order to operate on scattered nodes, we need to invoke some tools from meshless numerical analysis. Namely, we will employ Radial Basis Functions (RBFs) to approximate our unknowns. What follows is a brief description of the methods employed. A more in-depth explanation can be found in [6,27].

We start by introducing RBF interpolation. We are looking for an approximation of a function $f : \Omega \to \mathbb{R}$ that has known values $f_i = f(\mathbf{x}_i)$ on a scattered node set $X = \{\mathbf{x}_i\}_{i=1}^N$. The $N$ radial basis functions are defined by our discretisation as: $\Phi_i(\mathbf{x}) = \varphi(\|\mathbf{x} - \mathbf{x}_i\|)$ for some chosen radial basic function $\varphi$. The RBF interpolant is then:

$$s(\mathbf{x}) = \sum_{i=1}^{N} \alpha_i \Phi_i(\mathbf{x}). \tag{7}$$

Usually, $\varphi$ is chosen to be a strictly positive definite function (as are many popular choices, such as Gaussians or multiquadrics). In this case, the interpolation problem can be shown to admit a unique solution [6]. In practice, to achieve a certain degree of polynomial reproduction we augment our interpolant with monomials:

$$s(\mathbf{x}) = \sum_{i=1}^{N} \alpha_i \Phi_i(\mathbf{x}) + \sum_{j=1}^{M_m} \beta_j p_j(\mathbf{x}), \tag{8}$$

where $p_j(\mathbf{x})$ are monomials, and $M_m$ is the dimension of the space of all polynomials of degree up to $m$, inclusive.[4] The interpolant remains unique as long as the set $X$ is m-polynomially unisolvent [6], which is generally the case for irregular node layouts [28]). To get the unknown expansion coefficients $\alpha = (\alpha_1, \dots, \alpha_n)^T$ and $\beta = (\beta_1, \dots, \beta_{M_m})^T$ the following linear system must be solved:

$$\tilde{A} \begin{bmatrix} \alpha \\ \beta \end{bmatrix} = \begin{bmatrix} A & P \\ P^T & 0 \end{bmatrix} \begin{bmatrix} \alpha \\ \beta \end{bmatrix} = \begin{bmatrix} f \\ 0 \end{bmatrix}, \tag{9}$$

where $f = (f_1, \dots, f_n)^T$, $A_{ij} = \varphi(\|\mathbf{x}_i - \mathbf{x}_j\|)$ and $P_{ij} = p_j(\mathbf{x}_i)$. Defining $\Phi(\mathbf{x}) = (\Phi_1(\mathbf{x}), \dots, \Phi_N(\mathbf{x}))^T$ and $p(\mathbf{x}) = (p_1(\mathbf{x}), \dots, p_{M_m}(\mathbf{x}))^T$, we can rewrite the interpolant in a more compact form:

$$s(\mathbf{x}) = (\alpha \ \beta)^T (\Phi(\mathbf{x}) \ p(\mathbf{x})) = (f \ 0)^T \tilde{A}^{-1} (\Phi(\mathbf{x}) \ p(\mathbf{x})) = \sum_{i=1}^{N} \psi_i(\mathbf{x}) f_i, \tag{10}$$

where $\psi_i(x) = (\tilde{A}^{-1}(\Phi(\mathbf{x}) \ p(\mathbf{x}))_i$, is the $i$-th cardinal function[5] and satisfies $\psi_i(\mathbf{x}_j) = \delta_{ij}$.

So far, we have described a global method – the interpolation problem was solved over the whole domain. In practice, number of nodes $N$ is large and this quickly becomes computationally inefficient and ill-conditioned. For that reason, we are usually interested in local approximations near a given point $\mathbf{x}_c$ and we write the interpolant (8) over a certain neighbourhood of the point $\mathbf{x}_c$, commonly known as its stencil. A stencil is usually taken to simply consist of $n$ ($n \ll N$) nodes that are closest to $\mathbf{x}_c$ – an example can be seen on Fig. 3. If we denote the index set of the stencil as $S(\mathbf{x}_c)$, the local interpolant becomes:

$$s_c(\mathbf{x}) = \sum_{i\in S(\mathbf{x}_c)} \alpha_i^c \Phi_i(\mathbf{x}) + \sum_{j=1}^{M_m} \beta_j^c p_j(\mathbf{x}) = \sum_{i\in S(\mathbf{x}_c)} \psi_i^c(\mathbf{x}) f_i, \tag{11}$$

where the unknown coefficients depend on the center point $\mathbf{x}_c$ and can be obtained by solving a linear system, as before. It can be shown that for a sufficiently smooth $f$ the pointwise interpolation error scales as $\mathcal{O}(h^{m+1})$ [11].

[4] $M_m = \binom{m+d}{d}$, where $d$ is the dimensionality of the domain.
[5] Also known as Lagrange function.

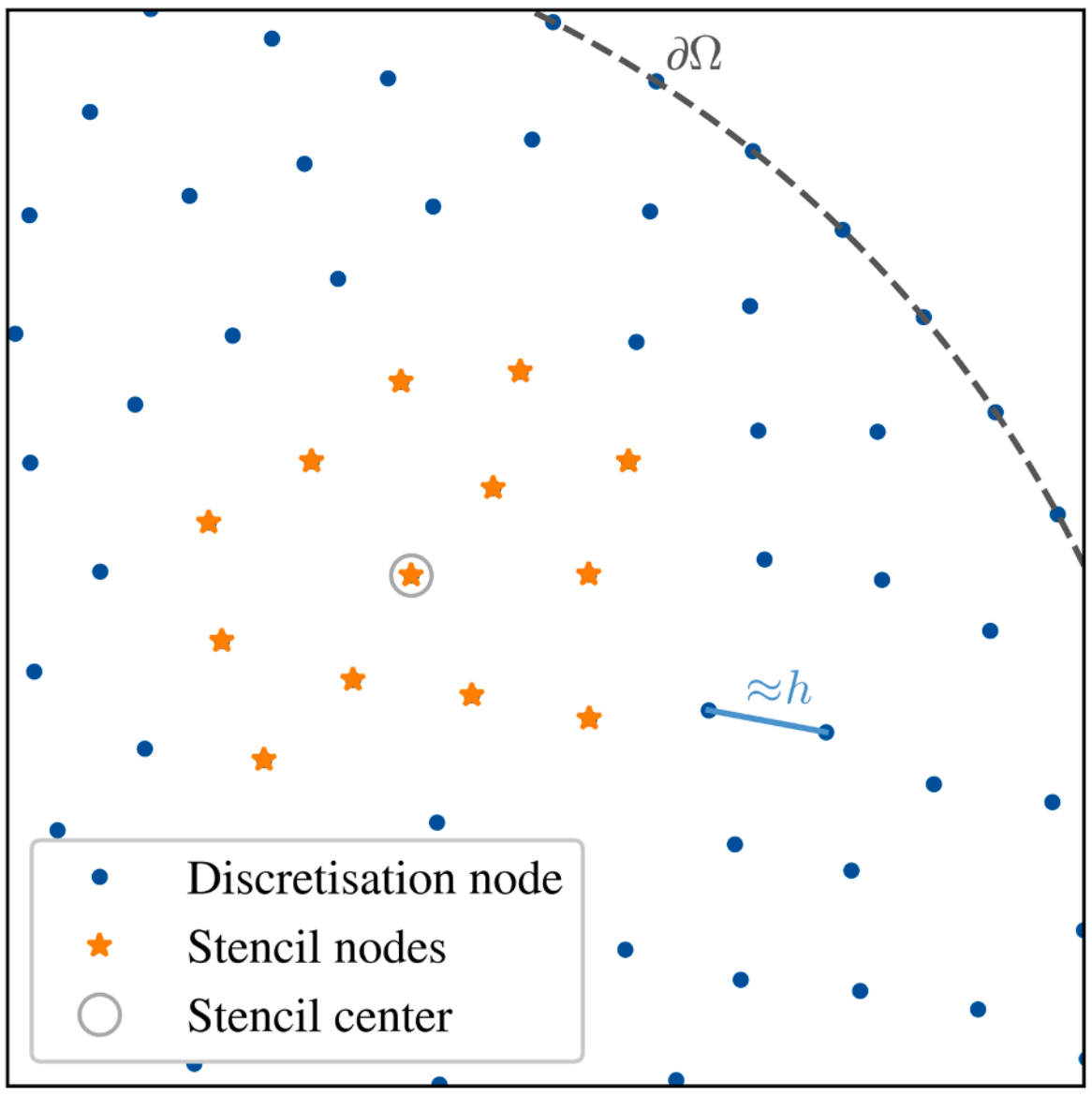

**Fig. 3.** An example scattered discretisation, with the nodal spacing of approximately $h$. The domain boundary $\partial\Omega$ and the stencil for a chosen node are also displayed.

### 4.2. Differential operator approximation

We aim to approximate a linear differential operator $\mathcal{L}$ as a linear combination of the function values in a stencil, akin to the usual finite differences:

$$\mathcal{L} f(\mathbf{x}_c) = \sum_{i\in S(\mathbf{x}_c)} w_i^{\mathcal{L}} f(\mathbf{x}_i) \tag{12}$$

We can obtain such an approximation using a local RBF interpolant in two different ways. The usual approach is the Radial Basis Function-generated Finite Differences (RBF-FD), which simply amounts to applying the operator $\mathcal{L}$ to the interpolant:

$$\mathcal{L} f(x_c) \approx \mathcal{L} s_c(x_c) = \sum_{i\in S(\mathbf{x}_c)} \mathcal{L}(\psi_i^c(\mathbf{x}_c)) f(\mathbf{x}_i) = \sum_{i\in S(\mathbf{x}_c)} w_i^{\mathcal{L}} f(\mathbf{x}_i). \tag{13}$$

The truncation error of this approach can be shown to scale as $\mathcal{O}(h^{m+1-l})$ [11], where $l$ is the order of the highest derivatives appearing in $\mathcal{L}$.

The second approach is not as widely used in the meshless community, although it is also based on RBF interpolantion and has been sucessfully applied to non-trivial problems [29,30]. It requires us to have a way of approximating $\mathcal{L}$ with the usual finite differences – we have a finite difference stencil given by coefficients $c = (c_1, \dots, c_k)$ and offsets $o = (\mathbf{o}_1, \dots, \mathbf{o}_k)$. For example, for the central derivative in the $x$ direction we have $c = (0, -1, 1)$ and $o = ((0,0), (-1,0), (1,0))$, while for the 5-point Laplace operator approximation we have $c = (-4, 1, 1, 1, 1)$, $o = ((0,0), (1,0), (-1,0), (0,1), (0,-1))$. Approximating the operator $\mathcal{L}$ in the finite difference manner then amounts to:

$$\mathcal{L} f(\mathbf{x}_c) \approx \delta^{-l} \sum_{i=1}^{k} c_i f(\mathbf{x}_c + \delta \mathbf{o}_i), \tag{14}$$

where $l$ is the order of the operator $\mathcal{L}$ and $\delta$ the spacing between the nodes. While this works well on a grid, the problem is that in a scattered layout $\mathbf{x}_c + \delta\mathbf{o}_i$ is in general not in our discretisation, so we do not have access to the value of $f$ at this point. We circumvent this by using the RBF interpolant instead, effectively imagining a virtual stencil around the point $x_c$, as seen on Fig. 4:

$$\mathcal{L}f(\mathbf{x}_c) \approx \delta^{-l} \sum_{i=1}^{k} c_i s_c(\mathbf{x}_c + \delta \mathbf{o}_i) = \delta^{-l} \sum_{i=1}^{k} c_i \sum_{j \in S(\mathbf{x}_c)} \psi_j^c(\mathbf{x}_c + \delta \mathbf{o}_i) f(\mathbf{x}_j) = \sum_{j \in S(\mathbf{x}_c)} w_j^{\mathcal{L}} f(\mathbf{x}_j), \tag{15}$$

where $w_j^{\mathcal{L}} = \delta^{-l} \sum_{i=1}^{k} c_i \psi_j^c(\mathbf{x}_c + \delta \mathbf{o}_i)$, are the differentiation weights expressing the approximation of the differential operator $\mathcal{L}$ and can again be understood as the generalised finite difference coefficients. The virtual stencil spacing $\delta$ is a parameter of the method and usually expressed in the units of the internodal spacing $h$ as $\delta = \sigma h$. Experiments show that it is desirable to have $\sigma < 1$ [29,31]. As far as the authors know, this second approach does not have a well-established name and we will refer to it as Radial Basis Function-generated Virtual Finite Differences (RBF-VFD) in this paper. The order of the RBF-VFD approximant can be easily deduced: error of the interpolation step scales as $\mathcal{O}(h^{m+1})$ and the interpolant is divided by $h^l$, reducing the exponent to $\mathcal{O}(h^{m+1-l})$. The second step, the application of the chosen finite difference stencil has a fixed, known order **ord**. The total order is then simply $\min(m+1-l, \mathbf{ord})$.

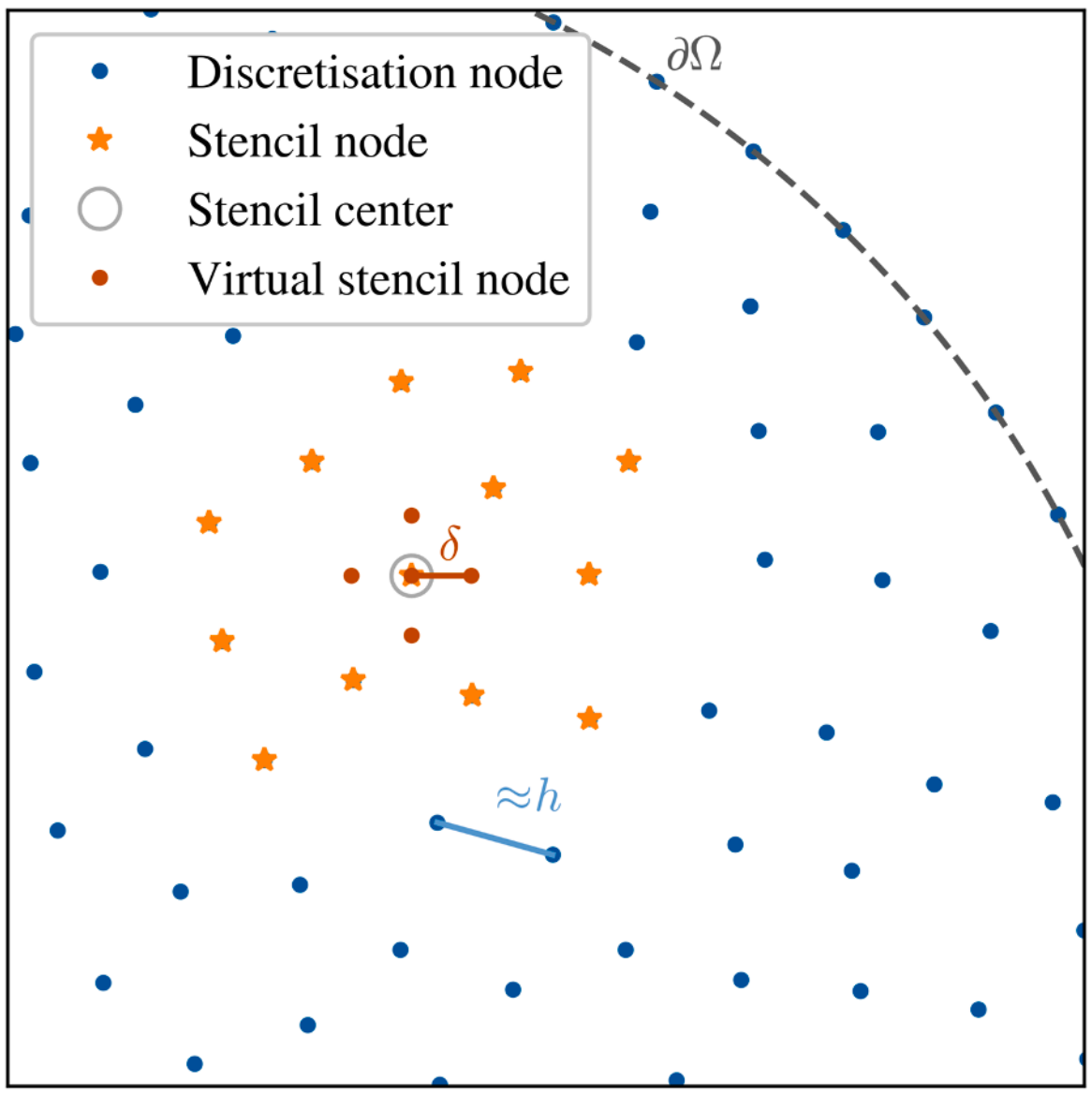

**Fig. 4.** A visual demonstration of the virtual stencil approach for the case of a 5-point stencil with spacing $\delta$.

### *4.3. Details of our setup*

In our work, we have exclusively used Polyharmonic Splines (PHS) as our chosen RBF. Concretely, we use radial cubics $\varphi(r) = r^3$, which are conditionally positive definite of order 2 and thus require the addition of monomials up to at least $m = 1$ [6]. It has been shown that such a combination can maintain the approximation power of monomials, while offering increased stability if the stencil size is chosen to be at least $n = 2M_m$ [12], which will be our default choice as well.

For meshless numerical analysis, we use an open source C++ library Medusa [32]. All of our source codes and data are publicly available on a git repository.[6]

## 5. Generalising the FDTD to a meshless setting

Having the tools of numerical analysis on scattered nodes behind us, we can now proceed and implement a variant of the FDTD in a meshless setting.

Firstly, the previously mentioned shortcomings of FDTD arise due to a regular discretisation in space. Time direction has trivial geometry and poses no issues, so we will keep the staggering of fields in time and the same time evolution method as in the FDTD – the interleaved leapfrog.

The problematic part is the spatial discretisation, which will be given by scattered nodes. As mentioned in the introduction, we have decided to discard the idea of staggering the $\vec{H}$ and $\vec{E}$ fields at different points. Instead, we simply take the FDTD update Eq. (5)

[6] https://gitlab.com/e62Lab/2025_p_meshless_computational_electromagnetics

and replace the spatial derivatives (central differences) by their suitable meshless substitutes:

$$\begin{aligned} H_x(x_i, y_i, n+1) &= H_x(x_i, y_i, n) - \frac{\Delta t}{\mu_0} D^{\partial_y}(E_z(x_i, y_i, n)), \\ H_y(x_i, y_i, n+1) &= H_y(x_i, y_i, n) + \frac{\Delta t}{\mu_0} D^{\partial_x}(E_z(x_i, y_i, n)), \\ E_z(x_i, y_i, n+1) &= E_z(x_i, y_i, n) + \frac{\Delta t}{\epsilon_0} \left(D^{\partial_x} H_y(x_i, y_i, n+1) - D^{\partial_y} H_x(x_i, y_i, n+1)\right), \end{aligned} \tag{16}$$

where we have introduced the notation $D^{\mathcal{L}} f(\mathbf{x}_i) = \sum_{j \in S(\mathbf{x}_i)} w_j^{\mathcal{L}} f(\mathbf{x}_j)$ and the two approaches that we will consider differ in how the differentiation weights $w_j^{\mathcal{L}}$ are calculated.

In the first approach, we will calculate the differentiation weights using RBF-VFD, where the finite difference stencils in question are simply the central differences. This approach seems to be as close to the Yee grid as we can get to in a meshless setting. We have opted for $\delta = 0.5h$ as our virtual stencil spacing of choice, which is within the previously mentioned guidelines and mirrors the spacing of the original Yee grid, where the $\vec{E}$ and $\vec{H}$ are also offset by the same amount. In the second approach we will apply RBF-FD to obtain the differentiation weights, which directly generalises the usual FDTD [33]. In order to match the order of FDTD, we calculate the derivative weights using $m = 2$ in both approaches.

The above scheme remains explicit and thus efficient, as the differentiation weights are only computed once at the beginning of the simulation and can be reused. We gain additional flexibility of the method due to the meshless nature of the differential operators. However, our attempt at a more geometrically flexible version of the FDTD comes at a cost in terms of the method stability. As mentioned, in FDTD it can be shown that the method is stable if the Courant number satisfies $S_c := \frac{c_0 \Delta t}{\Delta s} \leq S_c^{\max}$, where $S_c^{\max}$ equals $\frac{1}{\sqrt{2}}$ for the two-dimensional case with similar stability conditions available also for other regular node layouts [21]. However, in a meshless setting, the nodes are scattered and such an estimate is not easily reachable. We have attempted to stabilise the method by computing at an increasingly lower $S_c$ with no success,[7] hinting at an inherent instability of the method (refer to Appendix A for an illustration of why instability may be expected). To remedy these problems and make our proposed methods useful, a stabilisation procedure is needed.

### 5.1. Stabilising the method

Instabilities in explicit numerical schemes are a well-known problem and subject to a plethora of research in the scientific community. Several different approaches to stabilisation have been proposed over the years, such as penalty schemes [34] or upwinding [35], to name a few.

We have opted for an approach that has been gaining popularity in the meshless community lately: Hyperviscosity (HV) [36]. It can be effective and is in principle very easy to implement, as it amounts to simply adding a higher order derivative term to our system:

$$\begin{aligned} \frac{\partial H_x}{\partial t} &= -\frac{1}{\mu_0} \frac{\partial E_z}{\partial y} + \gamma_1 \Delta^{\alpha_1} H_x, \\ \frac{\partial H_y}{\partial t} &= \frac{1}{\mu_0} \frac{\partial E_z}{\partial x} + \gamma_2 \Delta^{\alpha_2} H_y, \\ \frac{\partial E_z}{\partial t} &= \frac{1}{\epsilon_0} \left( \frac{\partial H_y}{\partial x} - \frac{\partial H_x}{\partial y} \right) + \gamma_3 \Delta^{\alpha_3} E_z, \end{aligned} \tag{17}$$

where $\Delta^\alpha$ is an iterated Laplace operator. For further intuition on how HV stabilises a given method, we refer an interested reader to [36,37].

What remains is to determine parameters $\gamma_i, \alpha_i$, which should be such that the HV term stabilises the solution, while minimally affecting the physically relevant solution components. We first perform some simplifications that turned out to work well. We set:

$$\begin{aligned} \alpha_1 &= \alpha_2 = \alpha_3 = \alpha, \\ \gamma_1 &= \gamma_2 = \frac{\gamma}{\mu_0}, \\ \gamma_3 &= \frac{\gamma}{\epsilon_0}. \end{aligned} \tag{18}$$

These choices are motivated by the fact that we would like to have as few parameters as possible, while still taking into account the asymmetry between $\vec{H}$ and $\vec{E}$.

[7] For the forthcoming analyses, unless otherwise stated, we work with $S_c = (10\sqrt{2})^{-1}$.

Additionally, we follow the hyperviscosity guidelines described in [37] and set $\gamma = c\,(-1)^{\alpha-1}h^{2\alpha}$, for some $c > 0$. We will discretise the newly introduced hyperviscosity operator $\Delta^\alpha$ using RBF-FD, resulting in modified update equations of the stabilised method:

$$
\begin{aligned}
H_x(x_i, y_i, n+1) &= H_x(x_i, y_i, n) - \frac{\Delta t}{\mu_0}\big(D^{\partial_y}(E_z(x_i, y_i, n)) - \gamma D^{\Delta^\alpha} H_x(x_i, y_i, n)\big),\\
H_y(x_i, y_i, n+1) &= H_y(x_i, y_i, n) + \frac{\Delta t}{\mu_0}\big(D^{\partial_x}(E_z(x_i, y_i, n)) + \gamma D^{\Delta^\alpha} H_y(x_i, y_i, n)\big),\\
E_z(x_i, y_i, n+1) &= E_z(x_i, y_i, n) + \frac{\Delta t}{\epsilon_0}\big(D^{\partial_x} H_y(x_i, y_i, n+1) - D^{\partial_y} H_x(x_i, y_i, n+1) +\\
&\quad + \gamma D^{\Delta^\alpha} E_z(x_i, y_i, n)\big).
\end{aligned}
\tag{19}
$$

Note that, for simplicity, RBF-FD is used for HV terms even if the $\partial_x, \partial_y$ operators are discretised using RBF-VFD. The rationale behind this decision is that HV terms are merely a stabilisation procedure and should not be responsible for other properties of the method. Following also the guidelines presented in [38], we choose $\varphi_{\mathrm{HV}}(r) = r^{2\alpha+1}$, $m_{\mathrm{HV}} = 2\alpha$ and $n_{\mathrm{HV}} = 2M_{m_{\mathrm{HV}}} + 1$ for the RBF-FD discretisation of the HV operator.

The values of parameters $\alpha$ and $c$ are determined by a parameter sweep – we solve a given problem with some chosen parameters and observe its stability properties.

To quantify (in)stability of a method, we track the behaviour of energy over time for different values of $c$ and $\alpha$. The formula of energy of the electromagnetic field is known:

$$
\begin{aligned}
U(t) &\propto \int_\Omega \|\vec{E}(t)\|^2 + \eta_0^2\|\vec{H}(t)\|^2 \mathrm{d}\Omega \approx\\
&\approx h^2 \sum_{i=1}^{N} \big(E_z(x_i, y_i, t)^2 + \eta_0^2(H_x(x_i, y_i, t)^2 + H_y(x_i, y_i, t)^2)\big),
\end{aligned}
\tag{20}
$$

where the integral was approximated with a sum similarly as in [11]. Due to staggering in time, we do not have access to $\vec{E}$ and $\vec{H}$ fields at the same time points. In order to calculate the energy at an integer multiple of $\Delta t$, we must average the $\vec{H}$ fields (which are otherwise only defined at odd multiples of $\Delta t/2$). In practice, we track the relative energy $u(t) = U(t)/U(0)$, which should remain approximately 1 for a conservative problem setup. We should emphasize that no information on analytical solution of the problem is needed to compute $u(t)$.

As an example, we solve Maxwell's equations on a domain $\Omega = [0, 200]^2$, with periodic boundary conditions (emulating an unbounded domain) and initial condition given by:

$$
\begin{aligned}
E_z(x, y, t = 0) &= e^{-\frac{(x-\mu)^2}{2\cdot\sigma^2}},\\
H_x(x, y, t = -0.5\Delta t) &= 0,\\
H_y(x, y, t = -0.5\Delta t) &= -e^{-\frac{(x-\mu+0.5c_0\Delta t)^2}{2\cdot\sigma^2}}/\eta_0,
\end{aligned}
\tag{21}
$$

which is a Gaussian pulse, that should analytically move at a speed $c_0$ in the $x$ direction without changing its shape. We choose $\sigma = 30$ as its width and $\mu = 100$ as its offset.

The time dependence of $u(t)$ for different stabilisation parameters is displayed on Fig. 5, where the problem was solved with RBF-VFD. We can observe that $c$ has a major effect on the stability of the method and that a well-chosen value of $c$ and $\alpha$ leads to an approximate conservation of energy throughout the simulation. As a reference, the energy of the FDTD simulation with the same parameters (where applicable) is displayed. FDTD has no stability issues, resulting in $u(t) \approx 1$. Note that some difference in energy values between FDTD and RBF-VFD is expected – our crude integration scheme from Eq. (20) is more accurate on grids, where the distance between the two closest nodes is exactly $h$. On the other hand, in scattered nodes the distance between two closest nodes varies and is only approximately equal to $h$. On Fig. 5 we have only plotted a small number of hyperviscosity parameter combinations for clarity. A table displaying more combinations can be found in Appendix B.

A more complete picture of the effect of $c$ on stability can also be seen on Fig. 6, where relative energy after 10000 timesteps with respect to $c$ is shown. Here, a common pattern in HV can be observed: There exists an interval of suitable values of $c$, for which the solution is stable. For $c$ outside this interval, the solution diverges. Note that stable schemes may also be dissipative, which in our case happens only for $\alpha = 1$ and is undesirable.

For the remainder of this section, we have settled on $\alpha = 2$ and $c = 10^{-4}$ as it is a stable combination with little dissipation. Generally, we prefer a lower $\alpha$ if possible, as the stencil size and therefore the computational complexity of the method grows with $\alpha$.

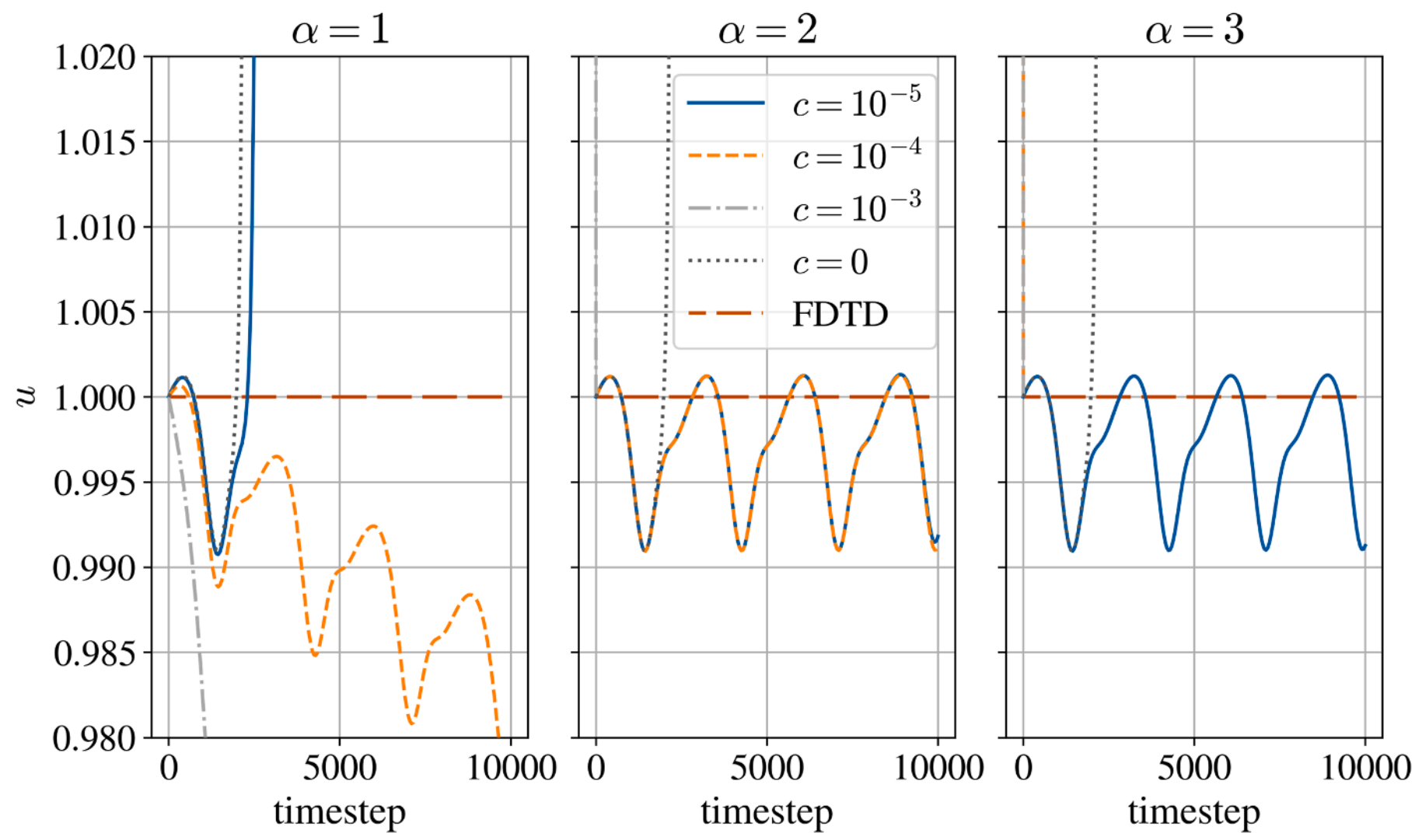

**Fig. 5.** Behaviour of relative energy $u$ over time for the considered test case for different hyperviscosity parameters $\alpha$ and $c$. The result obtained with the FDTD is also shown as a reference.

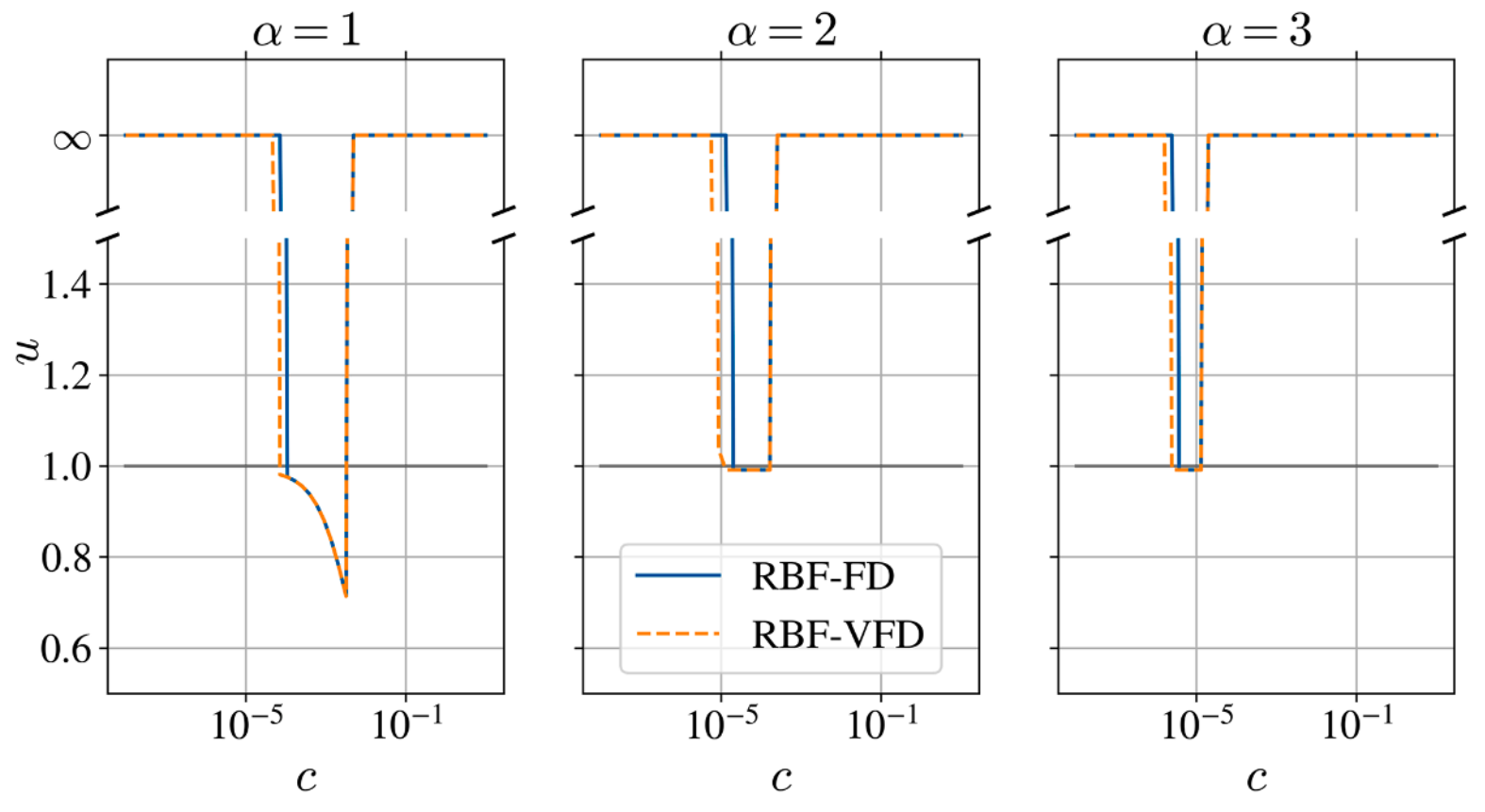

**Fig. 6.** Relative energy after 10000 timesteps with respect to $c$ for different values of the parameter $\alpha$. The value $\infty$ represents the divergence of the solution.

## *5.2. Convergence of the solution*

To end this section, let us verify that the proposed scheme is convergent. In the above setup $H_x, H_y$ were chosen such that the wave propagates into the positive $x$ direction (the propagation direction is given by $\vec{E} \times \vec{H}$). Since the problem is independent of $y$, it is effectively 1-dimensional and as $E_z$ satisfies the wave equation, the solution is given by the d'Alembert formula and equals:

$$\begin{aligned} &E_z^a(x, y, t) = E_z(x - c_0 t, y, 0) \\ &H_x^a(x, y, t) = 0, \\ &H_y^a(x, y, t) = -E_z(x - c_0 t, y, 0)/\eta_0, \end{aligned} \tag{22}$$

where we have added the superscript $a$ to denote the analytical solution and differentiate it from the numerically obtained solution. This allows us to verify that our stabilised method is indeed convergent. We check this by reducing discretisation distances in space and time, while keeping the Courant number (this time set to $S_c^{-1} = 100\sqrt{2}$) constant. To study convergence we will define the following two $\ell^\infty$ errors:

$$\begin{aligned} \|e\|_\infty^E &:= \max_{i=1}^N \left( \|\vec{E}(\mathbf{x}_i) - \vec{E}^a(\mathbf{x}_i)\|_\infty \right), \\ \|e\|_\infty^H &:= \max_{i=1}^N \left( \|\vec{H}(\mathbf{x}_i) - \vec{H}^a(\mathbf{x}_i)\|_\infty \right). \end{aligned} \tag{23}$$

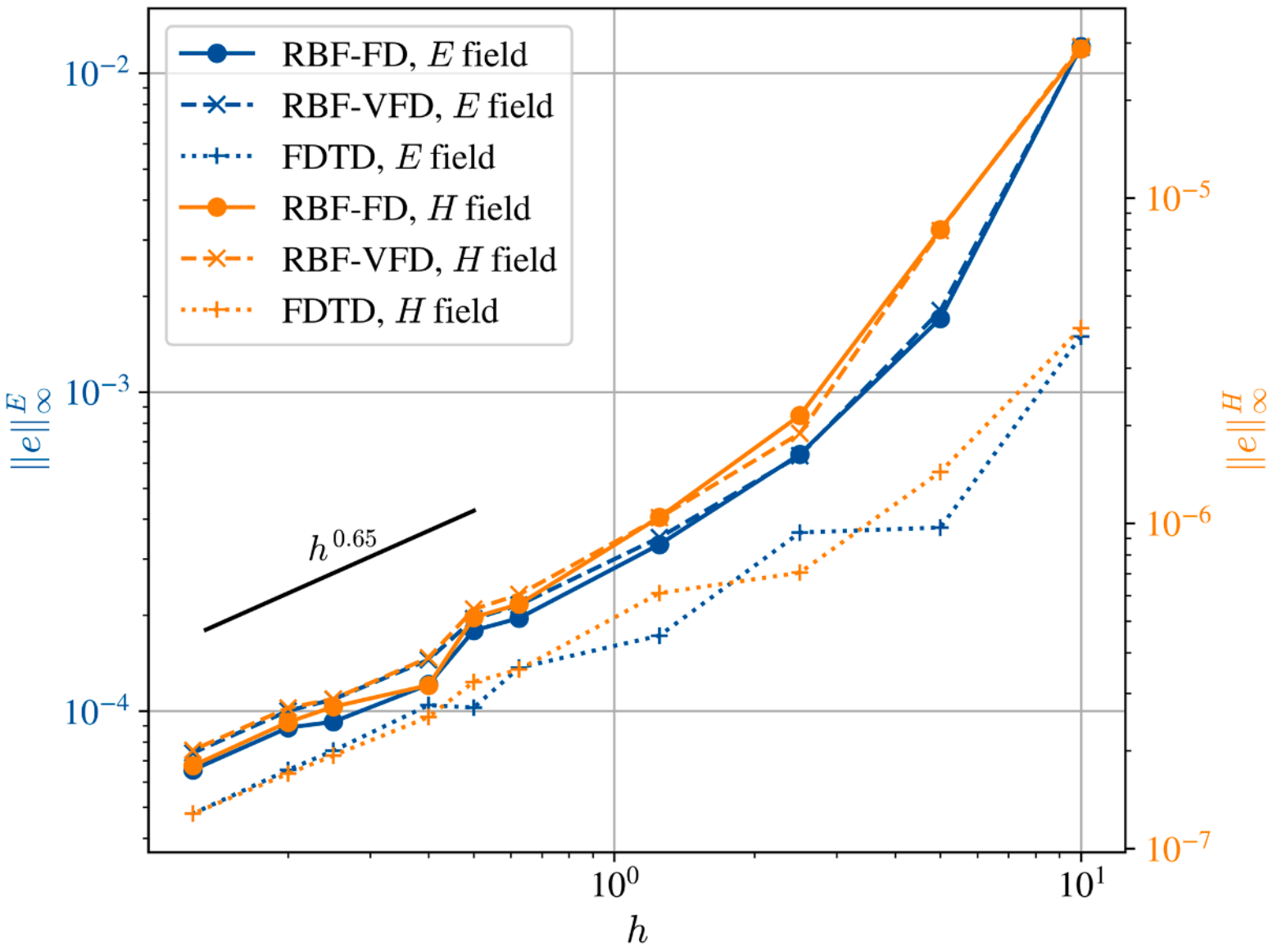

**Fig. 7.** $\|e\|_\infty^H$ and $\|e\|_\infty^E$ errors with respect to the discretisation distance $h$. Errors were computed at time $c_0 t = 5\sqrt{2}$.

Results are displayed in Fig. 7, where we can see that both methods are convergent with similar error behaviour. For FDTD, an $\mathcal{O}(h)$ convergence rate is expected for sufficiently small $h \to 0$. In our proposed methods a lower convergence rate would not be unexpected due to the presence of hyperviscosity.

To summarise, we now have a stable method that can operate on scattered nodes. With this we have addressed the first shortcoming of the FDTD and we now turn to the second - anisotropic dispersion relation.

## 6. Dispersion

Having developed a working meshless algorithm, we now turn to its dispersion properties. To evaluate the magnitude of dispersion of our methods, we consider a similar test case as in Section 5.1, except that we use $\sigma = 2$ - a very narrow Gaussian that is difficult to resolve at $h = 1$ and will push the method to its limits inducing visible effects of dispersion. As the setup is now different, the previously calculated HV parameters are not suitable and we recompute them. Due to the complexity of the new case we no longer get extended stable and non-dissipative regions we have observed on Fig. 6. Instead, a new hyperviscosity parameter sweep can be seen on Fig. 8, where we can see that dissipation now occurs even in $\alpha > 1$ case. An ideal hyperviscosity parameter $c$ in this case is the smallest stable $c$, denoted by a dash-dotted grey line, as it exhibits the least amount of dissipation, similar to the findings of [39]. In principle, such a $c$ could be determined with a very fine parameter sweep. However, it is much more efficient to use bisection, which is also explained in [39]. Using the bisection algorithm and $\alpha = 2$ we get $c \approx 10^{-4.77}$ for the RBF-FD and $c \approx 10^{-4.87}$ for RBF-VFD. This combination maintained $u(t) \in [0.99, 1.01]$ even after 10000 timesteps. The hyperviscosity parameters are now fine-tuned to greater precision compared to our previous tests, which is not unexpected, as we have already established that the considered test case is considerably more difficult to resolve. This precision could be relaxed, however, depending on how much dissipation we are willing to tolerate in a given time horizon.

To analyse dispersion we will turn to Fourier analysis, which works well on uniformly spaced nodes. For that reason, we perform an extra step when discretising $\Omega = [0, 200]^2$ - first, we discretise the line $y = 100$ with equispaced nodes $\{\mathbf{x}_i^e\}_i$, at a distance of $h = 1$ apart. These are then used as seed nodes in the same discretisation algorithm to obtain a complete discretisation of $\Omega$, effectively resulting in a scattered node layout, where some points are guaranteed to be equispaced on a line.

Fig. 9 shows the snapshot of the RBF-FD simulation after 2820 timesteps, which is approximately equal to one passage of the wavefront across the whole domain. We can observe that increasing the stencil size[8] improves the dispersion properties of the solution. In fact, for this particular case, the smallest recommended stencil size $n = 13$ contains numerical artefacts in the solution. We do not yet fully understand the reasoning behind this behaviour, although some ideas on how to analyse it will be mentioned in the conclusion. On Fig. 10 the snapshots are shown also for the RBF-VFD simulation and besides the previous observations, we can see that, perhaps surprisingly, dispersion seems much more prominent in the RBF-VFD approach compared to the RBF-FD and remains considerable even at larger stencils.

[8] Note that this refers to the stencil size of the original parts of the update equations. For HV terms, we continue to adhere to the previous guidelines.

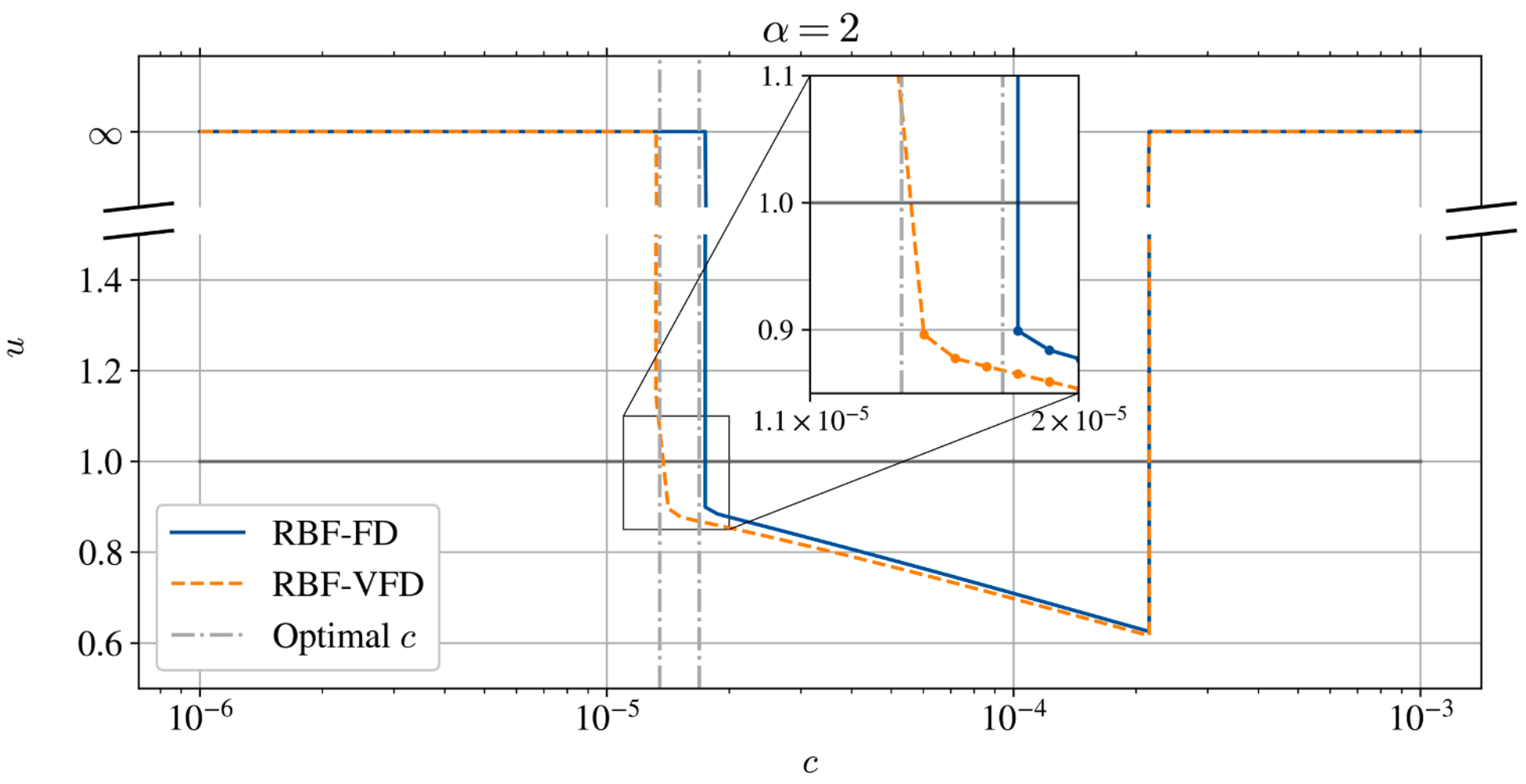

**Fig. 8.** Relative energy after 10000 timesteps with respect to $c$ for the $\sigma = 2$ case and $\alpha = 2$. Vertical, dash-dotted grey lines denote the optimal values of $c$ for the two methods as found by the bisection algorithm. The value $\infty$ represents the divergence of the solution.

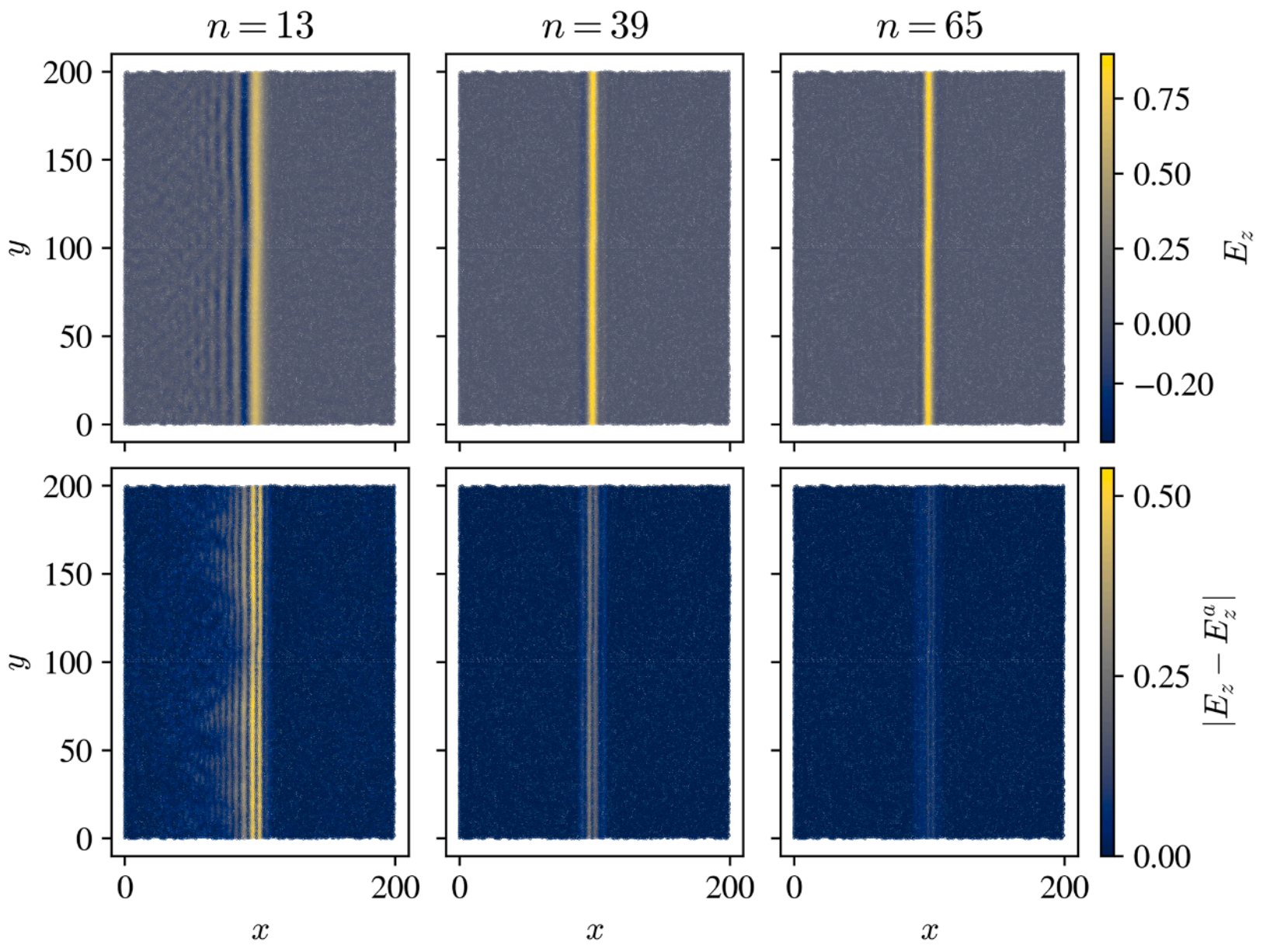

**Fig. 9.** Snapshot of the solution (upper row) and errors (lower row) obtained by RBF-FD at $c_0 t = 282/\sqrt{2}$ for different stencil sizes.

To put our observations in a quantitative setting, let us analyse the spectral picture of the solution. Running the simulation as before, we sample the solution values $E_z(\mathbf{x}_i^e, n\Delta t)$ and perform a two-dimensional Fourier transform to obtain the corresponding values in frequency space $E_z(k, \omega)$, where the allowed discrete values for $k$ are $k_i = \frac{2\pi i}{N_x}$, for $0 \le i < N_x$, with $N_x = 200$ (number of points on the discretised line) and similarly $\omega_j = \frac{2\pi j}{15 N_t}$ with $0 \le j < N_t$ with $N_t = 188$ (running the simulation for 2820 timesteps and saving every 15th snapshot).

Before transforming, Hann window was applied in the time direction to prevent spectral leakage due to non-periodicity (in time) of the solutions. The results can be seen in Fig. 11, where the absolute value of the individual spectral components $E_z(k_i, \omega_j)$ is displayed. Physically, we would expect all the modes to propagate at the same speed, that is, that the magnitude of $E_z(k, \omega)$ is large only when $\omega = c_0 k$, where $c_0 = \frac{1}{10\sqrt{2}}$ in our units. A black, dashed line representing this physical dispersion relation is also displayed in the same

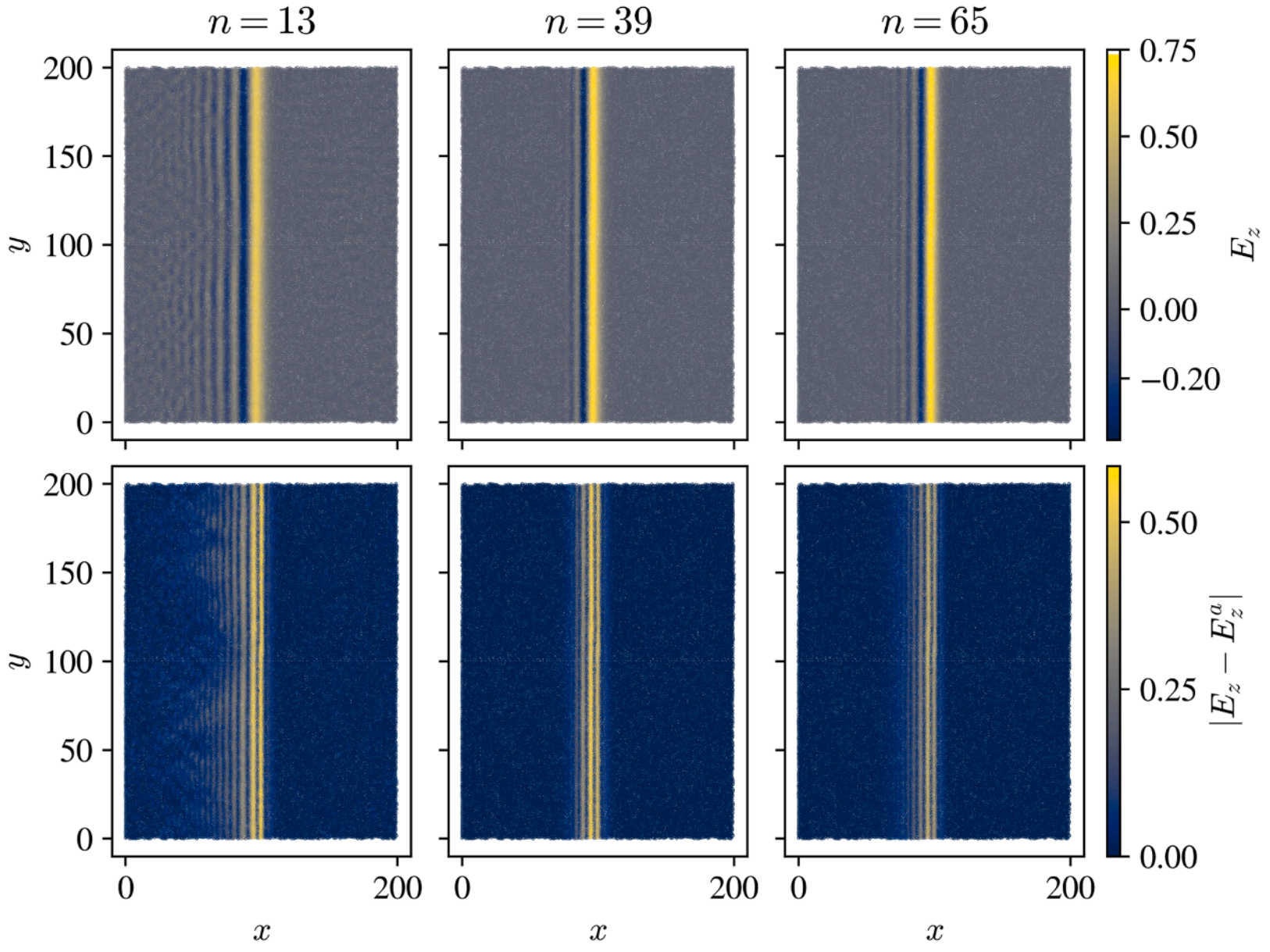

**Fig. 10.** Snapshot of the solution (upper row) and errors (lower row) obtained by RBF-VFD at $c_0 t = 282/\sqrt{2}$ for different stencil sizes.

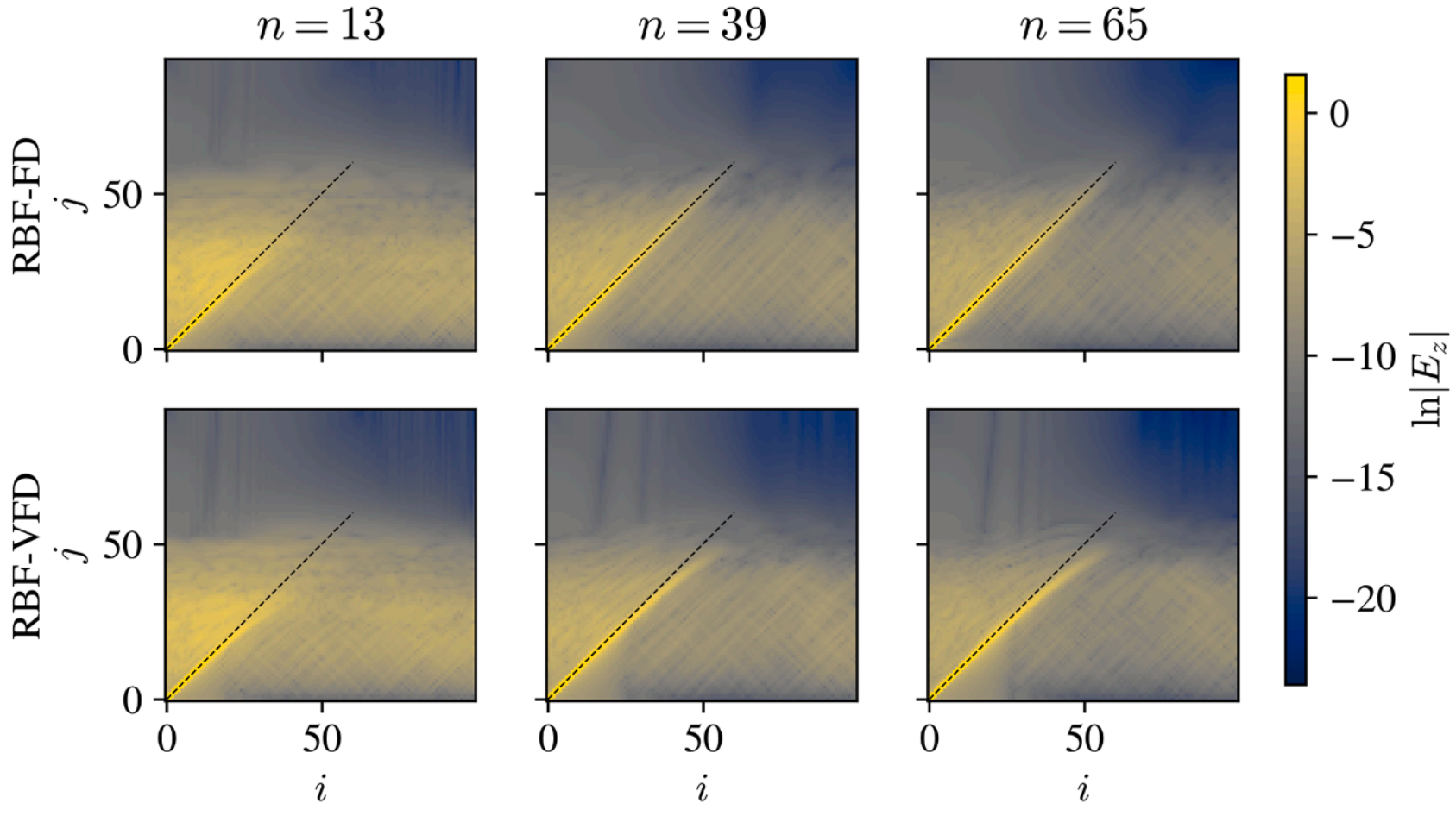

**Fig. 11.** $\ln|E_z(k,\omega)|$ for the considered cases. The $x$ and $y$ axes correspond to wavenumber and frequency, as indicated in the text. The expected dispersion relation is denoted by the black, dashed line.

Figure. We can clearly see the different behaviour of the cases with the smallest stencil size ($n = 13$), where the linear dispersion relation breaks down already at a relatively low frequency. This is greatly improved as stencil size is increased (note the logarithmic scale – the off-diagonal components are several orders of magnitude smaller). As already hinted by our previous simulation, we can observe that the RBF-VFD approach further deviates from the expected dispersion relation compared to the RBF-FD.

For the purposes of visualisation, we would like to summarise the amount of numerical dispersion with a single quantity. We first define a numerical propagation velocity $c(\omega_j)$ by selecting for each discrete frequency $\omega_j$ the corresponding wave vector $k_i$:

$$
\begin{aligned}
i(j) &:= \operatorname{argmax}_{0\le i<N_x} |E(k_i,\omega_j)|, \\
c(\omega_j) &:= \frac{\omega_j}{k_{i(j)}}.
\end{aligned}
\tag{24}
$$

The absolute error in the propagation velocity is then

$$
\mathbf{err}(\omega_j) := \left| c(\omega_j) - \frac{1}{10\sqrt{2}} \right|. \tag{25}
$$

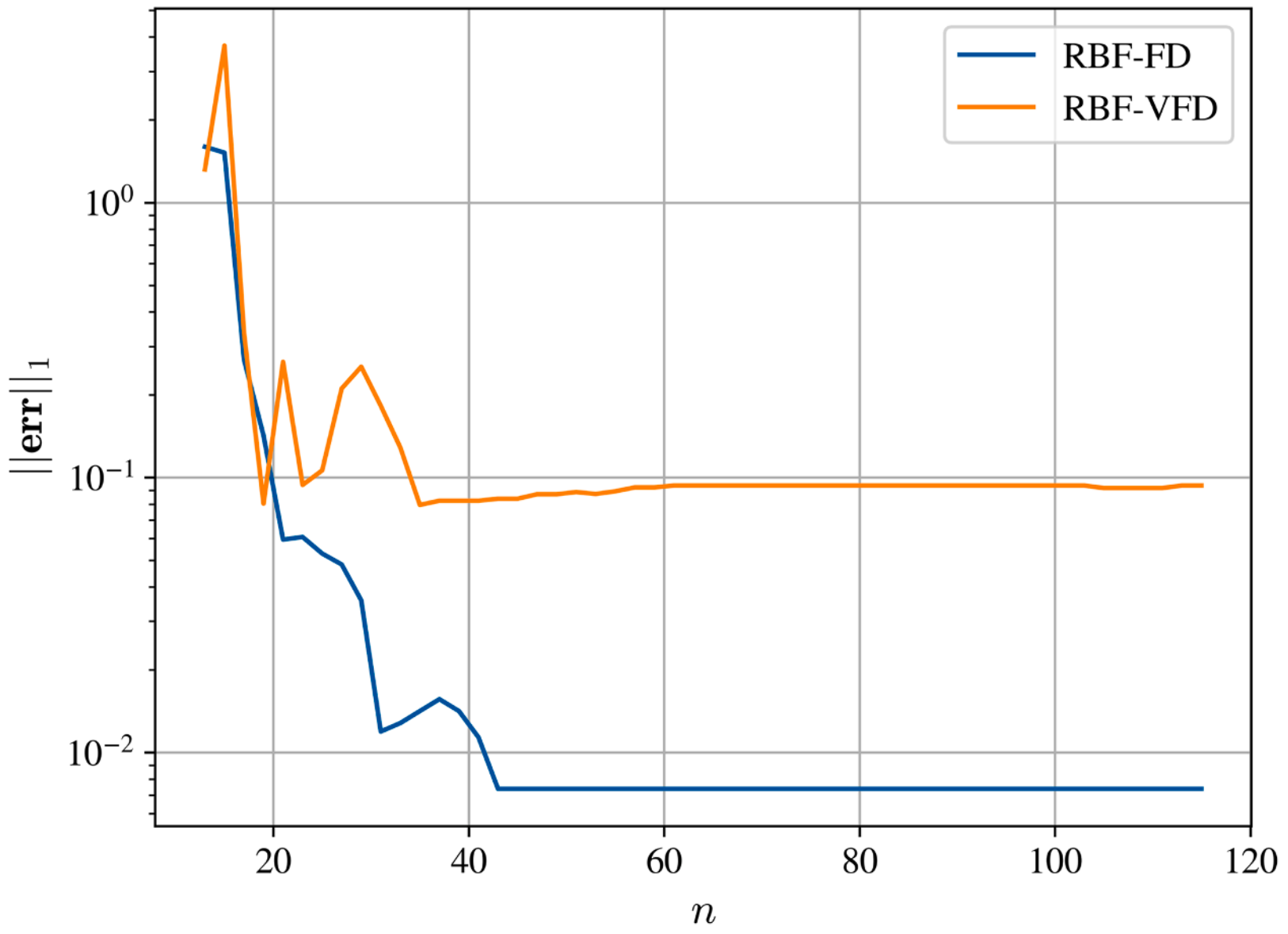

**Fig. 12.** Stencil size dependence of numerical dispersion.

As a quantity that summarises the amount of deviation from the physical numerical dispersion relation, we simply take the sligthly modified $\ell^1$ norm of $\mathbf{err}$:

$$\|\mathbf{err}\|_1 := \sum_{j=0}^{45} |\mathbf{err}(\omega_j)|, \tag{26}$$

where we have only summed up to $j = 45$, as higher values are not visibly present in the spectrum and would distort the results. This quantity is displayed in Fig. 12 with respect to the stencil size and matches our previous observations – numerical dispersion is large at small stencil sizes and is reduced as the stencil size is increased. From this point onward, we will work with $n = 60$ as our stencil size of choice for both methods.

Having delved into the dispersive properties of our presented methods it is time to introduce angular dependence and compare it to the FDTD. For this, we modify our setup yet again – we again start with $\Omega = [0, 200]^2$ and a discretised $y = 100$ line, rotate it by a chosen angle $\phi$, and afterwards complete the discretisation. The initial condition is effectively the same pulse wave as before, moving along the square of side $L$ (and rotated by $\phi$):

$$\begin{aligned}
&E_z(x, y, t = 0) = \exp(-(c_\phi x + s_\phi y - 0.5L)^2/(2\sigma^2)) \\
&H_x(x, y, t = -0.5\Delta t) = s_\phi \exp(-(c_\phi x + s_\phi y - 0.5L + 0.5c_0\Delta t)^2/(2\sigma^2))/\eta_0, \\
&H_y(x, y, t = -0.5\Delta t) = -c_\phi \exp(-(c_\phi x + s_\phi y - 0.5L + 0.5c_0\Delta t)^2/(2\sigma^2))/\eta_0,
\end{aligned} \tag{27}$$

where $s_\phi = \sin(\phi)$ and $c_\phi = \cos(\phi)$. Such a setup allows us to maintain periodic boundary conditions and perform the same Fourier analysis as before, the difference being that the pulse now propagates at an angle $\phi$ with respect to the $xy$ coordinate system.

Results of the simulation for two extreme angles $\phi = 0$ and $\phi = \frac{\pi}{4}$ are seen in Fig. 13, where the error snapshots are once again plotted. Note that, while for the $\phi = 0$ case, we can implement the FDTD on $[0, 200]^2$ as before, the rotated case was implemented on $[0, 144\sqrt{2}]^2$, rotated by $\frac{\pi}{4}$, as we cannot make the grid periodic otherwise. Qualitative results should be independent of the slight change in the domain size, as the width of the pulse remains the same. We can observe that the RBF-FD case is not affected by the orientation, which is what we initially hoped for, as the stencil has no directional bias. This directional independence was not a-priori obvious, however: In the RBF-FD approach we still compute the derivatives in the $x$ and $y$ direction separately and do not at any point account for the fact that the curl operator is rotationally invariant.

On the contrary, the dispersion properties of the FDTD are expectedly highly angle-dependent. Interestingly, the same holds for RBF-VFD, implying that having a stencil with no directional bias is not sufficient and the anisotropy can arise from the manner in which differentiation weights are computed.

We conclude the section by plotting $\|\mathbf{err}\|_1(\phi)$ in Fig. 14, where the RBF-FD and RBF-VFD curves were computed in the same manner as before, while for the FDTD we calculated $\mathbf{err}$ from Eq. (6). This plot again shows that the RBF-FD approach exhibits a much lower degree of dispersion anisotropy compared to the RBF-VFD and the FDTD, which behave similarly.

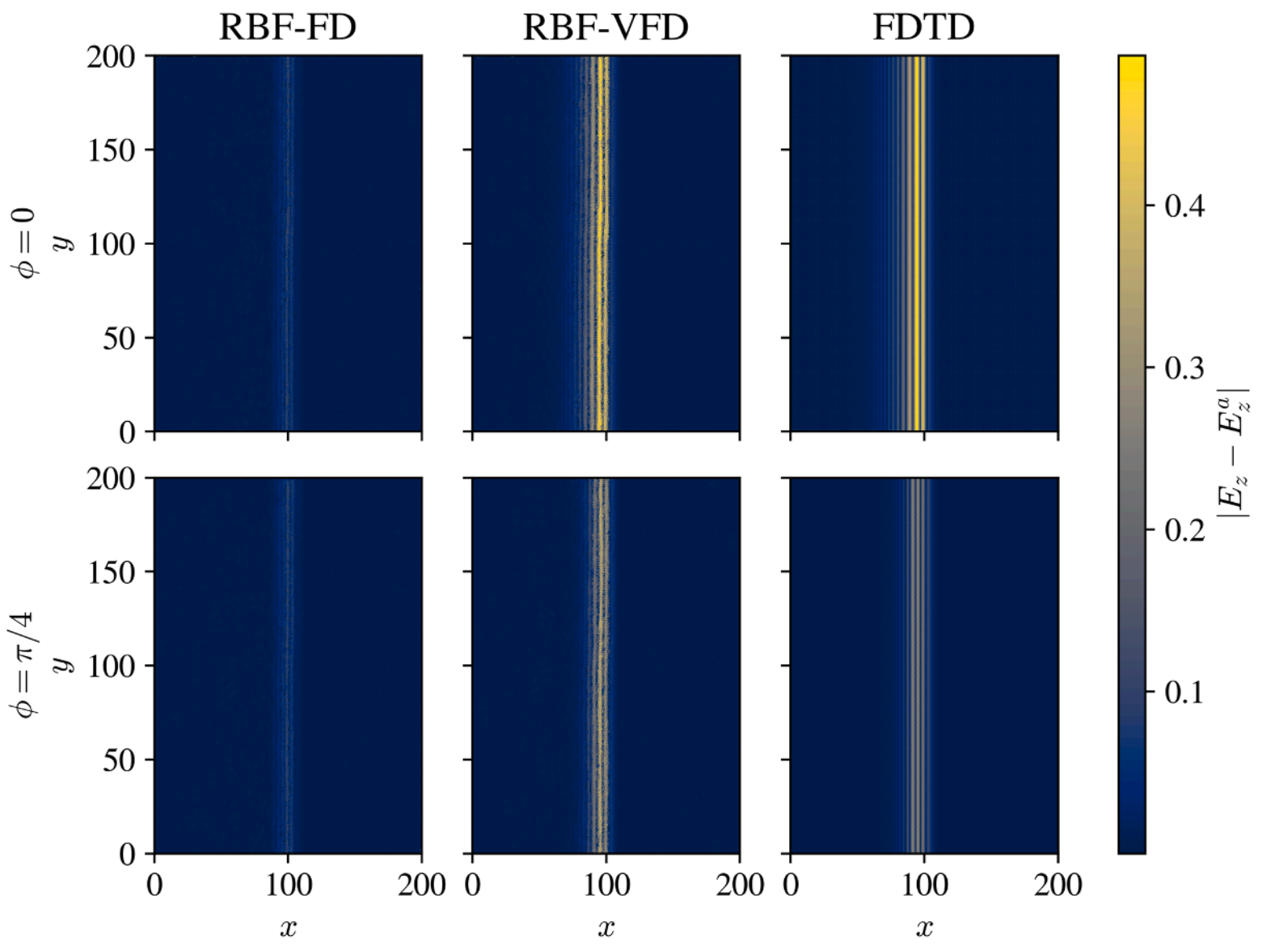

**Fig. 13.** Snapshots of the simulation errors after one passage through the domain for two extremal angles.

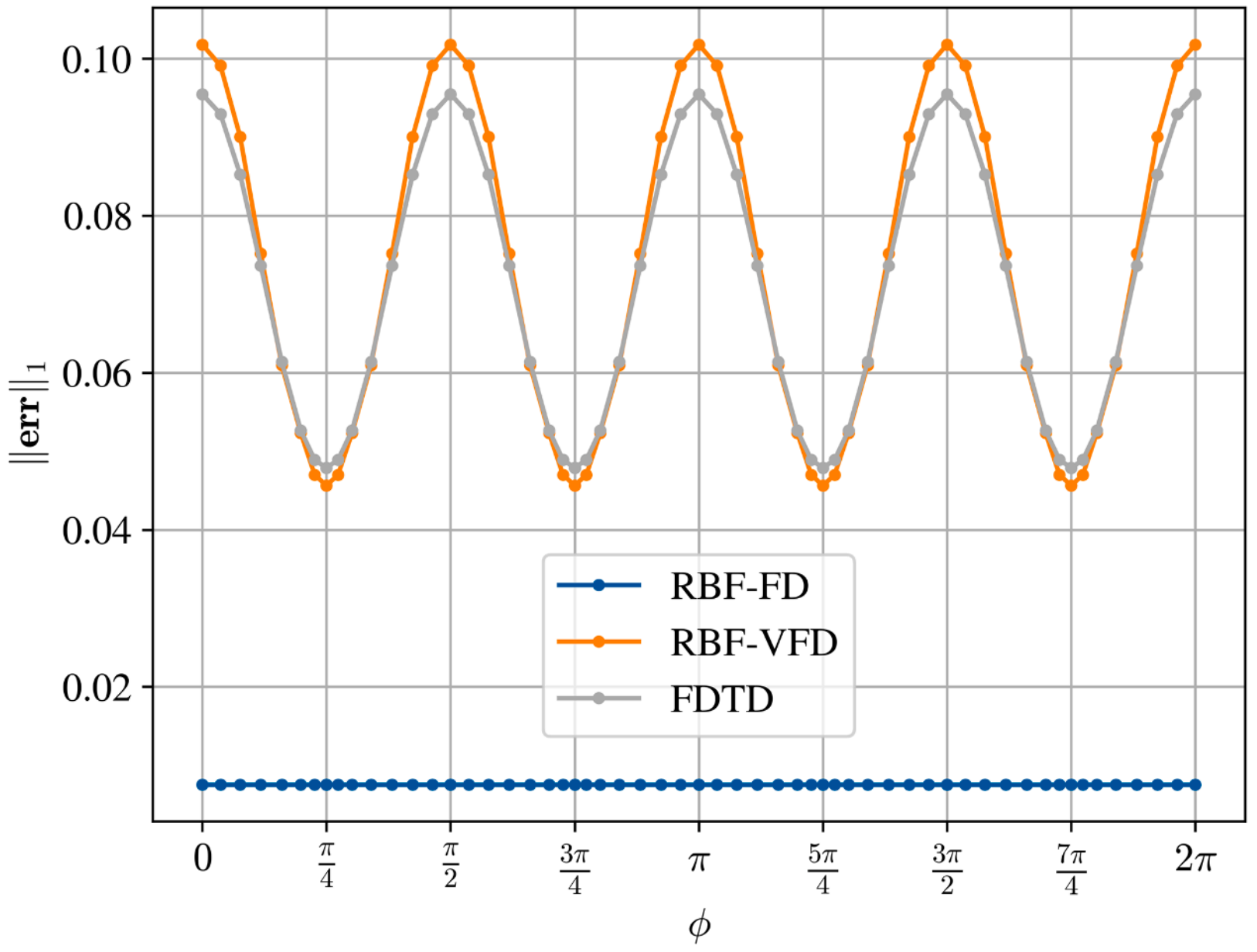

**Fig. 14.** Numerical dispersion dependence on the propagation angle $\phi$.

## 7. Moving towards scattering problems

Having developed a promising meshless generalisation of the FDTD, a natural next step is to apply it to scattering scenarios – a main subject of interest when considering time domain simulations of electromagnetic waves.

A fully-fledged meshless scattering framework requires an appropriate addition of several techniques commonly used in the FDTD besides our developed method [23]: To effectively introduce an incident wave, we should develop a meshless variant of the Total-Field Scattered-Field (TFSF), for instance. Periodic boundary conditions that we have used until this point are also not appropriate for scattering scenarios, and absorbing boundary conditions would have to be implemented. As the main contribution of this paper is the development of the proposed methods themselves, we leave the development of the full scattering framework as our future work.

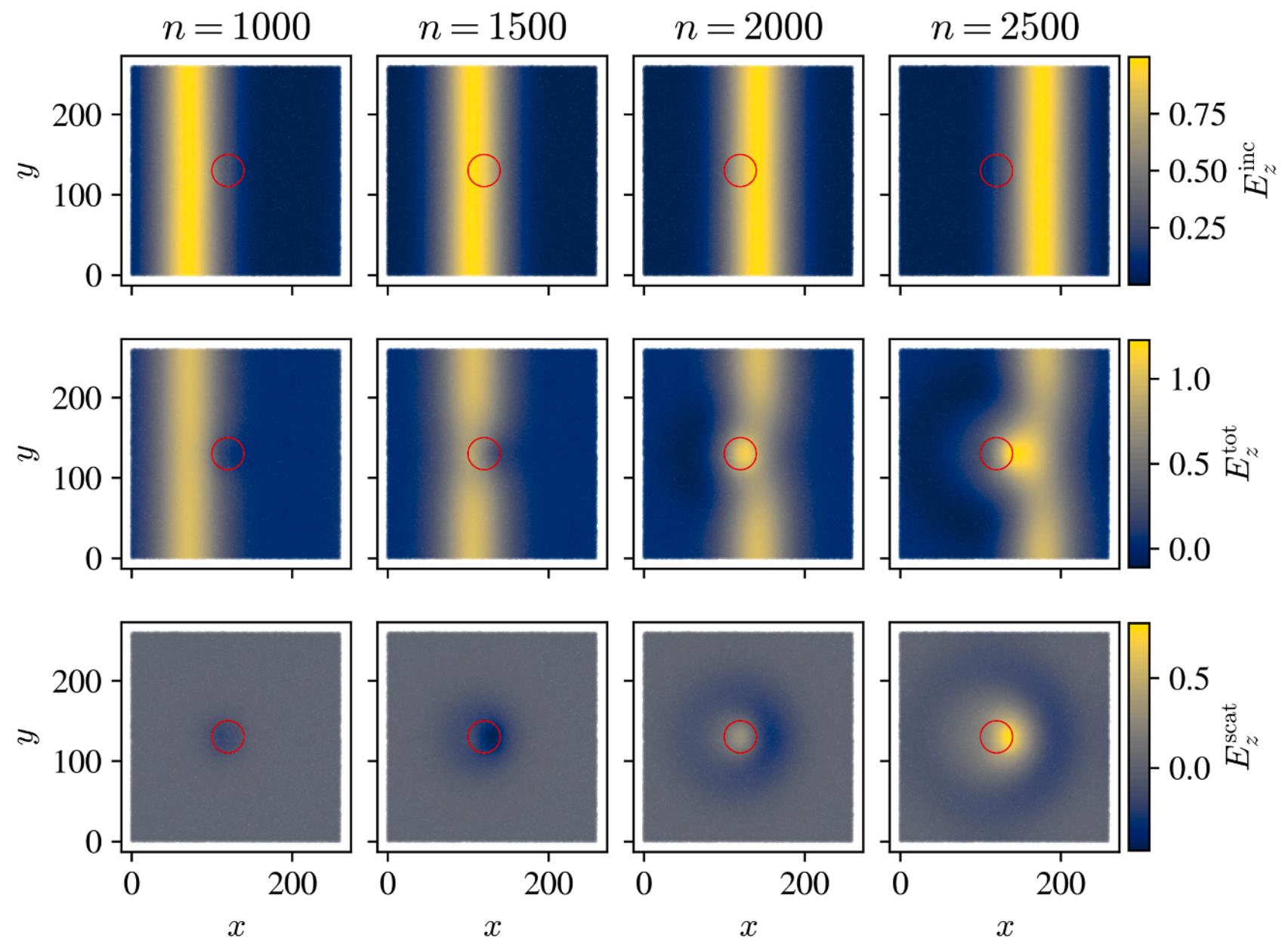

**Fig. 15.** Snapshots of the incident (first row), total (second row) and scattered (third row) fields obtained with the RBF-FD method. The red circle corresponds to the cylindrical scatterer with $\epsilon = 4$. (For interpretation of the references to colour in this figure legend, the reader is referred to the web version of this article.)

Still, as a proof of concept we will consider a simplified scattering test case. Our physical domain will now be $\Omega = [0, 260]^2$, except that we have a scatterer of cylindrical shape. Namely, our relative permittivity is now spatially dependent:

$$\epsilon_r(x, y) = \begin{cases} \epsilon, & \text{if } (x - 120)^2 + (y - 130)^2 \leq 20^2, \\ 1, & \text{otherwise.} \end{cases} \tag{28}$$

Our incident wave will be the Gaussian pulse that we have already considered in Eq. (27) with the same width of $\sigma = 30$ and an offset of $\mu = 0$. The incident pulse therefore starts far from the scatterer and continuously approaches it.

We also maintain the same value of the Courant number $S_c^{-1} = 10\sqrt{2}$ and hyperviscosity parameters $c = 10^{-4}$ and $\alpha = 2$ as in Section 5.1. We only increase the stencil size to $n = 60$ as we have since seen it is beneficial for dispersion. The simulation will be terminated at $c_0 t = 250/\sqrt{2}$, which is early enough to not feel the presence of the domain boundaries, circumventing the need for absorbing boundary conditions.

We will run two simulations with identical spatial and temporal discretisations. In the first one, we omit the scatterer and observe only the Gaussian pulse propagating through free space, as before. This gives us our incident wave $E_z^{\text{inc}}$. In the second simulation we modify the update Eq. (19), replacing all occurences of $\epsilon_0$ by $\epsilon_r(x, y)\epsilon_0$ to account for the scatterer. This gives us the total field $E_z^{\text{tot}}$.

We are most interested in the scattered field obtained from the two simulations as $E_z^{\text{scat}} = E_z^{\text{tot}} - E_z^{\text{inc}}$. Snapshots of $E_z^{\text{scat}}$ can be seen on Fig. 15, which has been obtained by our stabilised RBF-FD method on a spatial discretisation with $h = 1$ and a scatterer with $\epsilon = 4$. The first row contains only our incident pulse $E_z^{\text{inc}}$, while second and third row show the total $E_z^{\text{tot}}$ and scattered $E_z^{\text{scat}}$ fields, respectively. Results obtained by the FDTD are visually identical.

The considered test case of a Gaussian pulse is especially convenient for scattering, as it allows us to analyse the behaviour of several different frequencies in a single simulation – after Fourier transforming the involved fields we can calculate a specific quantity of interest, such as the scattering cross section, for each frequency mode present in the pulse. An efficient development of this machinery is part of the afromentioned meshless scattering framework.

Instead we can, with the tools that we have at our disposal, get a feel of how the results change for different frequencies by simply modifying the incident field to be a plane wave:

$$E_z^{\text{inc}} = \cos(kx), \tag{29}$$

where $k$ is a chosen wave number and $H_x^{\text{inc}}$, $H_y^{\text{inc}}$ are again chosen in a way to make the incident wave move in the $+x$ direction as in Eq. (22).

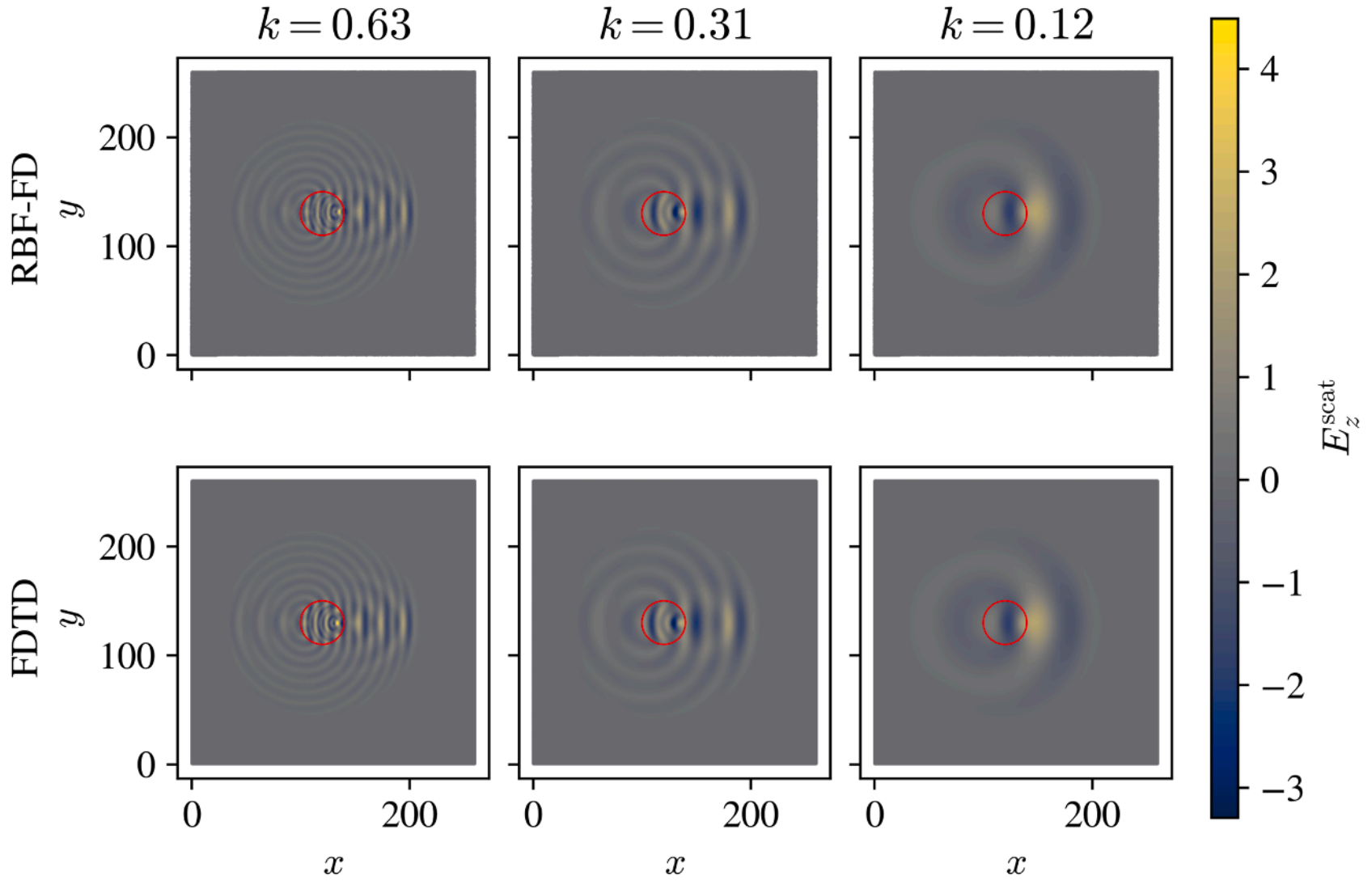

**Fig. 16.** Snapshots of the scattered field at $c_0 t = 100/\sqrt{2}$ for the RBF-FD (first row) and the FDTD (second row) methods for different values of wavenumber $k$. The red circle corresponds to the cylindrical scatterer with $\varepsilon = 4$. (For interpretation of the references to colour in this figure legend, the reader is referred to the web version of this article.)

Solution snapshots for a few different choices of $k$ are shown on Fig. 16, where we can see that the RBF-FD results are visually very similar to the FDTD results. The two simulations have been done at $h = 0.5$, again with a scatterer of $\varepsilon = 4$.

Fig. 16 shows that our proposed method can achieve comparable errors to FDTD also in the context of scattering. Note that, while these results appear optimistic, our current framework is not suitable for steady state scattering analyses, as illustrated in Appendix C. Nevertheless, the above results still provide an initial step towards further, practically applicable developments.

## 8. Conclusion

The starting point of our research was a seemingly simple question – how can the FDTD method be generalised to a scattered setting and would this help with the numerical dispersion anisotropy? The first question was answered affirmatively with the use of RBF-based meshless methods – we have performed a straightforward generalisation of the FDTD by simply replacing the derivative operators appropriately. We have considered two approaches, which a priori both seemed equally promising: The usual RBF-FD and the virtual stencil approach, in this paper referred to as the RBF-VFD. Unlike FDTD, which enjoys a simple stability criterion, major issues were found regarding stability of the proposed methods and we have demonstrated that these issues can be overcome by employing hyperviscosity stabilisation, describing a simple framework on how the appropriate HV parameters should be selected.

We then analysed the spectral picture of the solution and found that our proposed methods possess a parameter that can control the amount of dispersion - stencil size. Moving to the angular dependence of the dispersion, we have answered affirmatively also to the second part of our initial question - meshless methods can eliminate dispersion anisotropy, which in our example happened with the RBF-FD method. However, this is not always the case, as demonstrated by the RBF-VFD approach.

Ultimately, we have seen that RBF-FD performs better than FDTD in this sense and can be considered as a promising candidate for a meshless generalisation of the FDTD. We have further supported this statement by showing that we can reproduce the FDTD results also in a specific scattering scenario, marking the start of our work towards concrete, practical applications of the developed method.

Several important points were glossed over in our research that should be addressed if the proposed methods are ever to be used in an industrial setting. First, our analysis was for the most part limited to an effectively one-dimensional case and a very simple scattering scenario of a cylinder. As part of our current research, we are applying those same methods to tackle different cases, such as scattering on an irregular body or some 3D problems. This also includes analyses of the behaviour of the method when applied to cases with spatially varying discretisation density and the development of the complete framework for meshless scattering analysis. Additionally, considering approximations of higher order should be investigated, since this could also, as mentioned, further reduce numerical dispersion.

An important advantage of the FDTD method that was not given much attention is its intrinsic divergence conservation [21]. In this aspect, the RBF-VFD, being in some sense closer to the FDTD, could perform better compared to the RBF-FD approach, especially if paired with divergence-free RBFs [40], paving the way for potential future work in that direction.

Finally, the stencil dependence of numerical dispersion remains unexplained. We strongly suspect that the reason behind it can be connected to the recent developments in the RBF-FD method [11], where it was shown that the basis of our interpolant in the global sense is discontinuous, with the jumps decreasing with the stencil size. Further work on this matter could put our observations in a formal setting with supporting mathematical justifications.

## CRediT authorship contribution statement

**Andrej Kolar-Požun:** Writing – review & editing, Writing – original draft, Visualization, Validation, Software, Methodology, Investigation, Formal analysis, Conceptualization; **Gregor Kosec:** Writing – review & editing, Supervision, Resources, Funding acquisition, Conceptualization.

## Data Availability

The code is available on a public GitLab repository, which is linked in the paper.

## Declaration of competing interest

The authors declare that they have no known competing financial interests or personal relationships that could have appeared to influence the work reported in this paper.

## Acknowledgements

The authors would like to acknowledge the financial support from the Slovenian Research and Innovation Agency (ARIS) in the framework of the research core funding No. P2-0095, The Young Researcher program PR-12347 and research project No. J2-3048.

Furthermore, the authors would like to thank the following individuals for their assistance with various aspect of this study, in alphabetical order: Levstik Vito, Osmanov Vagif, Rogan Adrijan, Rot Miha and Vaupotič Žiga.

## Appendix A. Naive instability on scattered nodes

The purpose of this appendix is to provide some intuition on why a naive approach to solving Maxwell's equations on scattered nodes without any stabilisation fails.

For convenience, let us repeat the update equations from Section 5:

$$\begin{aligned}
H_x(x_i, y_i, n+1) &= H_x(x_i, y_i, n) - \frac{\Delta t}{\mu_0} D^{\partial_y}(E_z(x_i, y_i, n)),\\
H_y(x_i, y_i, n+1) &= H_y(x_i, y_i, n) + \frac{\Delta t}{\mu_0} D^{\partial_x}(E_z(x_i, y_i, n)),\\
E_z(x_i, y_i, n+1) &= E_z(x_i, y_i, n) + \frac{\Delta t}{\epsilon_0}\left(D^{\partial_x} H_y(x_i, y_i, n+1) - D^{\partial_y} H_x(x_i, y_i, n+1)\right),
\end{aligned} \tag{A.1}$$

where $D^{\mathcal{L}} f(x, y) = \sum_j w_j^{\mathcal{L}} f(x_j, y_j)$ is an approximation of a differential operator, given as a linear combination of function values in a stencil.

A straightforward way to analyse the stability of a method in FD schemes is by the means of von Neumann stability analysis [41]. Note that von Neumann analysis assumes translational invariance and hence applies to the usual grid-based schemes. In a meshless setting it should be used with extreme caution and not as a robust tool to assess the stability, perhaps providing only a very localised picture of modes that are amplified after a single iteration. Nevertheless, a variant of von Neumann analysis has been used by different authors in the meshless community with moderate success [37,42].

We take an ansatz:[9]

$$\begin{aligned}
H_x^n(x, y) &= H_{x0}^n e^{ik(x+y)}\\
H_y^n(x, y) &= H_{y0}^n e^{ik(x+y)}\\
E_z^n(x, y) &= E_{z0}^n e^{ik(x+y)}
\end{aligned} \tag{A.2}$$

[9] a general Fourier component has $i(k_x x + k_y y)$ in the exponential. However, we will see that unstable modes arise already in this special case.

Considering the update Eq. (A.1) for a single point $(x_i, y_i)$, plugging in the ansatz and dividing by $e^{ik(x+y)}$ we get:

$$\begin{aligned} H_{x0}^{n+1} &= H_{x0}^{n} - c_1 w_y E_{z0}^{n} \\ H_{y0}^{n+1} &= H_{y0}^{n} + c_1 w_x E_{z0}^{n} \\ E_{z0}^{n+1} &= E_{z0}^{n} + c_2\left(H_{y0}^{n+1} w_x - H_{x0}^{n+1} w_y\right) \end{aligned} \tag{A.3}$$

where we have introduced:

$$\begin{aligned} c_1 &:= \frac{\Delta t}{\mu_0} \\ c_2 &:= \frac{\Delta t}{\epsilon_0} \\ w_x &:= \sum_j w_j^{\partial_x} e^{ik(x_j - x + y_j - y)} \\ w_y &:= \sum_j w_j^{\partial_y} e^{ik(x_j - x + y_j - y)} \end{aligned} \tag{A.4}$$

Rewriting in matrix form, we get:

$$\begin{pmatrix} 1 & 0 & 0 \\ 0 & 1 & 0 \\ w_y c_2 & -w_x c_2 & 1 \end{pmatrix} \begin{pmatrix} H_{x0}^{n+1} \\ H_{y0}^{n+1} \\ E_{z0}^{n+1} \end{pmatrix} = \begin{pmatrix} 1 & 0 & -c_1 w_y \\ 0 & 1 & c_1 w_x \\ 0 & 0 & 1 \end{pmatrix} \begin{pmatrix} H_{x0}^{n} \\ H_{y0}^{n} \\ E_{z0}^{n} \end{pmatrix} \tag{A.5}$$

or, equivalently:

$$\begin{pmatrix} H_{x0}^{n+1} \\ H_{y0}^{n+1} \\ E_{z0}^{n+1} \end{pmatrix} = \begin{pmatrix} 1 & 0 & -c_1 w_y \\ 0 & 1 & c_1 w_x \\ -c_2 w_y & c_2 w_x & 1 + c_1 c_2 (w_x^2 + w_y^2) \end{pmatrix} \begin{pmatrix} H_{x0}^{n} \\ H_{y0}^{n} \\ E_{z0}^{n} \end{pmatrix} \tag{A.6}$$

For stability, it is necessary that the spectral radius of the matrix in the last equation does not exceed unity. We can calculate the eigenvalues to equal:

$$\begin{aligned} \lambda_1 &= 1 \\ \lambda_{2,3} &= 0.5\left(2 + c_1 c_2 (w_x^2 + w_y^2) \pm \sqrt{c_1 c_2}\sqrt{w_x^2 + w_y^2}\sqrt{4 + c_1 c_2 (w_x^2 + w_y^2)}\right) \end{aligned} \tag{A.7}$$

We can expect amplification if one of the $\lambda_{2,3}$ has modulus greater than unity. Denoting $\xi := c_1 c_2 (w_x^2 + w_y^2)$ we can write:

$$\lambda_{2,3} = 1 + 0.5\left(\xi \pm \sqrt{\xi^2 + 4\xi}\right) \tag{A.8}$$

Observe that the product of the two equals one:

$$\lambda_2 * \lambda_3 = (1 + 0.5\xi)^2 - 0.5^2(\xi^2 + 4\xi) = 1. \tag{A.9}$$

Therefore, $\lambda_2 = 1/\lambda_3$. From this it follows that if $|\lambda_2| < 1$ then $|\lambda_3| > 1$ or vice-versa and an unstable mode has been found. We must only exclude the possiblity of $|\lambda_2| = |\lambda_3| = 1$. However, in this case we can write $\lambda_2 = e^{i\alpha}$, $\lambda_3 = e^{-i\alpha}$ and calculate:

$$2 + \xi = \lambda_2 + \lambda_3 = e^{i\alpha} + e^{-i\alpha} = 2\cos(\alpha). \tag{A.10}$$

We notice that the case $|\lambda_2| = |\lambda_3| = 1$ is only possible if $\xi$ is real, which is something that is very unlikely to happen in an irregular stencil (real or imaginary parts in $w_x$,$w_y$ would have to exactly cancel out).

As emphasised earlier, this is by no means a formal proof of instability of any meshless approach, but instead merely an intuition as to where the locally unstable modes may come from. The analysis was limited to an interleaved leapfrog time-stepping, with no field staggering and a general expression of a derivative as a linear combination of stencil values. This setup includes our proposed RBF-FD and RBF-VFD based methods.

## Appendix B. Relative energies of additional hyperviscosity parameter combinations

This appendix is meant to supplement Fig. 5, where the time dependence of relative energy was shown for a few chosen combinations of hyperviscosity parameters. We present some further values of $u(\mathrm{timestep})$ for different parameters in tables below. Tables B.1–B.3 contain data for RBF-VFD, while Tables B.4–B.6 contain data for RBF-FD.

**Table B.1**
Parameter sweep results for RBF-VFD, $\alpha = 1$.

| $c$ | $u(0)$ | $u(500)$ | $u(1000)$ | $u(1500)$ | $u(2000)$ | $u(2500)$ | $u(3000)$ |
|---|---|---|---|---|---|---|---|
| $10^{-7}$ | 1.00 | 1.00 | 1.00 | 0.99 | 1.00 | 3.03 | 1184.32 |
| $10^{-6}$ | 1.00 | 1.00 | 1.00 | 0.99 | 1.00 | 2.34 | 711.10 |
| $10^{-5}$ | 1.00 | 1.00 | 1.00 | 0.99 | 1.00 | 1.02 | 5.92 |
| $10^{-4}$ | 1.00 | 1.00 | 1.00 | 0.99 | 0.99 | 0.99 | 1.00 |
| $10^{-3}$ | 1.00 | 0.99 | 0.98 | 0.97 | 0.97 | 0.96 | 0.96 |
| $10^{-2}$ | 1.00 | $\infty$ | $\infty$ | $\infty$ | $\infty$ | $\infty$ | $\infty$ |
| $10^{-1}$ | 1.00 | $\infty$ | $\infty$ | $\infty$ | $\infty$ | $\infty$ | $\infty$ |

**Table B.2**
Parameter sweep results for RBF-VFD, $\alpha = 2$.

| $c$ | $u(0)$ | $u(500)$ | $u(1000)$ | $u(1500)$ | $u(2000)$ | $u(2500)$ | $u(3000)$ |
|---|---|---|---|---|---|---|---|
| $10^{-7}$ | 1.00 | 1.00 | 1.00 | 0.99 | 1.00 | 2.41 | 754.40 |
| $10^{-6}$ | 1.00 | 1.00 | 1.00 | 0.99 | 1.00 | 1.04 | 9.90 |
| $10^{-5}$ | 1.00 | 1.00 | 1.00 | 0.99 | 1.00 | 1.00 | 1.00 |
| $10^{-4}$ | 1.00 | 1.00 | 1.00 | 0.99 | 1.00 | 1.00 | 1.00 |
| $10^{-3}$ | 1.00 | $\infty$ | $\infty$ | $\infty$ | $\infty$ | $\infty$ | $\infty$ |
| $10^{-2}$ | 1.00 | $\infty$ | $\infty$ | $\infty$ | $\infty$ | $\infty$ | $\infty$ |
| $10^{-1}$ | 1.00 | $\infty$ | $\infty$ | $\infty$ | $\infty$ | $\infty$ | $\infty$ |

**Table B.3**
Parameter sweep results for RBF-VFD, $\alpha = 3$.

| $c$ | $u(0)$ | $u(500)$ | $u(1000)$ | $u(1500)$ | $u(2000)$ | $u(2500)$ | $u(3000)$ |
|---|---|---|---|---|---|---|---|
| $10^{-7}$ | 1.00 | 1.00 | 1.00 | 0.99 | 1.00 | 1.07 | 16.91 |
| $10^{-6}$ | 1.00 | 1.00 | 1.00 | 0.99 | 1.00 | 1.00 | 1.00 |
| $10^{-5}$ | 1.00 | 1.00 | 1.00 | 0.99 | 1.00 | 1.00 | 1.00 |
| $10^{-4}$ | 1.00 | $\infty$ | $\infty$ | $\infty$ | $\infty$ | $\infty$ | $\infty$ |
| $10^{-3}$ | 1.00 | $\infty$ | $\infty$ | $\infty$ | $\infty$ | $\infty$ | $\infty$ |
| $10^{-2}$ | 1.00 | $\infty$ | $\infty$ | $\infty$ | $\infty$ | $\infty$ | $\infty$ |
| $10^{-1}$ | 1.00 | $\infty$ | $\infty$ | $\infty$ | $\infty$ | $\infty$ | $\infty$ |

**Table B.4**
Parameter sweep results for RBF-FD, $\alpha = 1$.

| $c$ | $u(0)$ | $u(500)$ | $u(1000)$ | $u(1500)$ | $u(2000)$ | $u(2500)$ | $u(3000)$ |
|---|---|---|---|---|---|---|---|
| $10^{-7}$ | 1.00 | 1.00 | 1.00 | 0.99 | 16.28 | 127509.84 | 1011952883.71 |
| $10^{-6}$ | 1.00 | 1.00 | 1.00 | 0.99 | 11.77 | 82565.22 | 601543100.26 |
| $10^{-5}$ | 1.00 | 1.00 | 1.00 | 0.99 | 1.35 | 1194.07 | 3795798.73 |
| $10^{-4}$ | 1.00 | 1.00 | 1.00 | 0.99 | 0.99 | 0.99 | 1.00 |
| $10^{-3}$ | 1.00 | 0.99 | 0.98 | 0.97 | 0.97 | 0.96 | 0.96 |
| $10^{-2}$ | 1.00 | $\infty$ | $\infty$ | $\infty$ | $\infty$ | $\infty$ | $\infty$ |
| $10^{-1}$ | 1.00 | $\infty$ | $\infty$ | $\infty$ | $\infty$ | $\infty$ | $\infty$ |

**Table B.5**
Parameter sweep results for RBF-FD, $\alpha = 2$.

| $c$ | $u(0)$ | $u(500)$ | $u(1000)$ | $u(1500)$ | $u(2000)$ | $u(2500)$ | $u(3000)$ |
|---|---|---|---|---|---|---|---|
| $10^{-7}$ | 1.00 | 1.00 | 1.00 | 0.99 | 12.21 | 86818.56 | 636991140.68 |
| $10^{-6}$ | 1.00 | 1.00 | 1.00 | 0.99 | 1.55 | 2055.42 | 7089752.71 |
| $10^{-5}$ | 1.00 | 1.00 | 1.00 | 0.99 | 1.00 | 1.00 | 1.00 |
| $10^{-4}$ | 1.00 | 1.00 | 1.00 | 0.99 | 1.00 | 1.00 | 1.00 |
| $10^{-3}$ | 1.00 | $\infty$ | $\infty$ | $\infty$ | $\infty$ | $\infty$ | $\infty$ |
| $10^{-2}$ | 1.00 | $\infty$ | $\infty$ | $\infty$ | $\infty$ | $\infty$ | $\infty$ |
| $10^{-1}$ | 1.00 | $\infty$ | $\infty$ | $\infty$ | $\infty$ | $\infty$ | $\infty$ |

**Table B.6**
Parameter sweep results for RBF-FD, $\alpha = 3$.

| $c$ | $u(0)$ | $u(500)$ | $u(1000)$ | $u(1500)$ | $u(2000)$ | $u(2500)$ | $u(3000)$ |
|---|---|---|---|---|---|---|---|
| $10^{-7}$ | 1.00 | 1.00 | 1.00 | 0.99 | 1.76 | 3053.05 | 11236149.17 |
| $10^{-6}$ | 1.00 | 1.00 | 1.00 | 0.99 | 1.00 | 1.00 | 1.00 |
| $10^{-5}$ | 1.00 | 1.00 | 1.00 | 0.99 | 1.00 | 1.00 | 1.00 |
| $10^{-4}$ | 1.00 | $\infty$ | $\infty$ | $\infty$ | $\infty$ | $\infty$ | $\infty$ |
| $10^{-3}$ | 1.00 | $\infty$ | $\infty$ | $\infty$ | $\infty$ | $\infty$ | $\infty$ |
| $10^{-2}$ | 1.00 | $\infty$ | $\infty$ | $\infty$ | $\infty$ | $\infty$ | $\infty$ |
| $10^{-1}$ | 1.00 | $\infty$ | $\infty$ | $\infty$ | $\infty$ | $\infty$ | $\infty$ |

## Appendix C. About the convergence of the scattering test case to the steady state solution

Towards the end of Section 7 we have seen that the RBF-FD and FDTD methods give very similar results when applied to the case of a plane wave scattering on a dielectric cylinder of relative permittivity $\varepsilon$.

The considered case actually has an analytical solution available [43], allowing us to analyse the convergence of the two methods also in this setup. In order to do so we compute the maximum error

$$\|\tilde{e}\|_{\infty}^{E} := \max_i \left(|E_z^{\text{scat}}(\mathbf{x}_i) - E_z^{a}(\mathbf{x}_i)|\right), \tag{C.1}$$

where $E_z^a$ is the analytical solution and the index $i$ goes over all the nodes in $[150, 160] \times [120, 140]$. This region was chosen as it is positioned in an area of interest, right behind the scatterer, while still being far enough from the domain boundaries. Our simulations were compared to the analytical solution at $c_0 t = 200/\sqrt{2}$ and performed with $\varepsilon = 1.77$. Other parameters remain the same as in Section 7.

The results can be seen on Fig. C.1. We can observe that initially the error decreases with decreasing $h$ (and therefore $\Delta t$), however the error curves start to flatten out as $h \to 0$. The reason for that is likely the fact that we have not yet reached the steady state and transient phenomena are still present in the domain, but not in the reference solution.

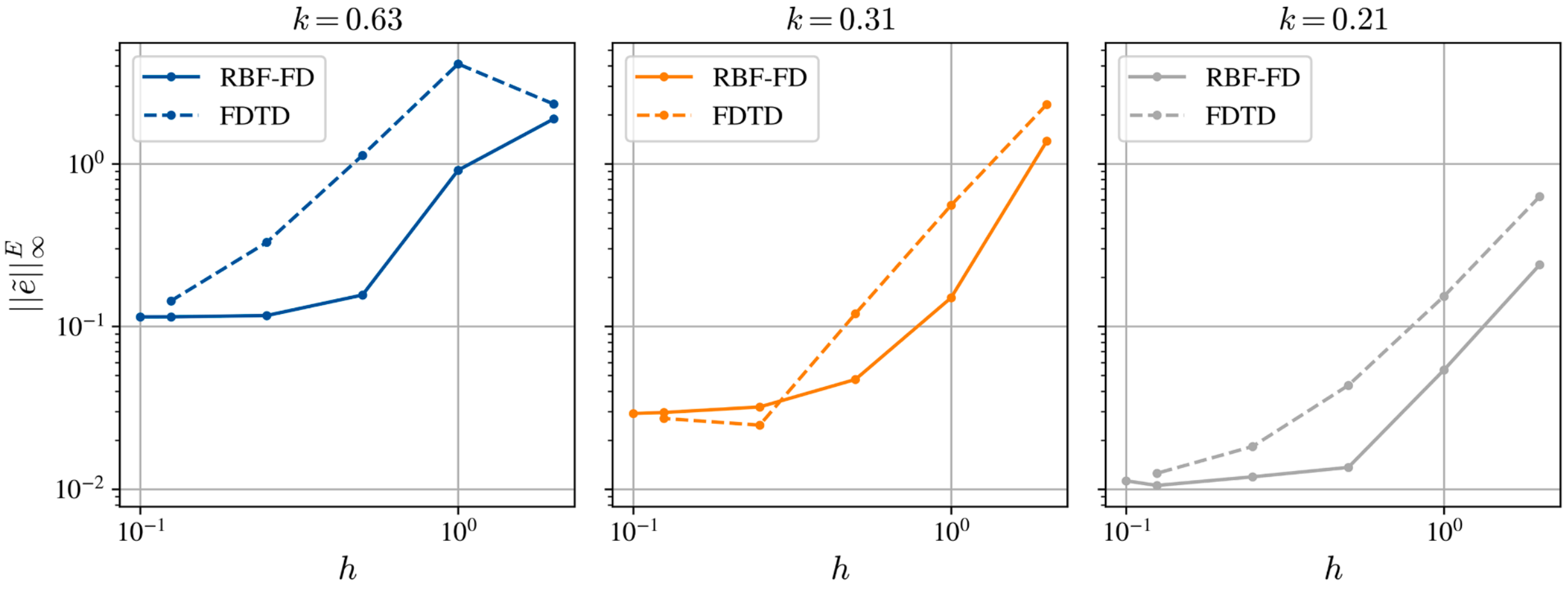

**Fig. C.1.** Convergence of the maximum absolute error at chosen points for the scattering example. The scatterer in this case has $\varepsilon = 1.77$.

Simply letting the simulation run for longer is not feasible within the current framework, which is the main limiting factor of studying scattering without the techniques mentioned at the beginning of Section 7. The main issue is that, due to a lack of absorbing boundary conditions, pushing to longer times gets computationally very expensive – we have to simultaneously enlarge the domain to prevent boundary effects and for the smallest $h$ considered in this study, the total number of points in the domain is already a few million.

Note that this is not an issue of RBF-FD per se – we observe the same behaviour also in FDTD. However, in FDTD the solution is readily available (absorbing boundary conditions), while in a meshless setting further modification may be needed.